\documentclass[reprint, aps, prl, twocolumn]{revtex4-1}
\usepackage{amsmath,amssymb,braket,fancybox}
\usepackage{amsfonts}
\usepackage{graphicx}
\usepackage{comment}
\usepackage{ascmac}
\usepackage{ascmac}
\usepackage{amsthm}
\usepackage{color}

\usepackage{lineno}
\newcommand*\patchAmsMathEnvironmentForLineno[1]{
  \expandafter\let\csname old#1\expandafter\endcsname\csname #1\endcsname
  \expandafter\let\csname oldend#1\expandafter\endcsname\csname end#1\endcsname
  \renewenvironment{#1}
     {\linenomath\csname old#1\endcsname}
     {\csname oldend#1\endcsname\endlinenomath}}
\newcommand*\patchBothAmsMathEnvironmentsForLineno[1]{
  \patchAmsMathEnvironmentForLineno{#1}
  \patchAmsMathEnvironmentForLineno{#1*}}
\AtBeginDocument{
\patchBothAmsMathEnvironmentsForLineno{equation}
\patchBothAmsMathEnvironmentsForLineno{align}
\patchBothAmsMathEnvironmentsForLineno{flalign}
\patchBothAmsMathEnvironmentsForLineno{alignat}
\patchBothAmsMathEnvironmentsForLineno{gather}
\patchBothAmsMathEnvironmentsForLineno{multline}
}

\theoremstyle{definition}

\newtheorem*{theorem*}{Theorem}

\newtheorem*{definition*}{Definition}

\newcommand{\mc}[1]{\mathcal{#1}}
\newcommand{\mr}[1]{\mathrm{#1}}
\newcommand{\mbb}[1]{\mathbb{#1}}

\newcommand{\lrs}[1]{\left( #1 \right)}

\newcommand{\lrl}[1]{\left[ #1 \right]}
\newcommand{\lrv}[1]{\left| #1 \right|}

\newcommand{\fracd}[2]{\frac{\mathrm{d} #1 }{\mathrm{d} #2 }}

\newcommand{\aln}[1]{
\begin{align}
#1
\end{align}
}

\newcommand{\ra}{\rightarrow}

\newcommand{\Tr}{\mr{Tr}}

\begin{document}
\title{
Exceptional Dynamical Quantum Phase Transitions in Periodically Driven Systems
}
\date{\today}
\author{Ryusuke Hamazaki}
\affiliation{
Nonequilibrium Quantum Statistical Mechanics RIKEN Hakubi Research Team, RIKEN Cluster for Pioneering Research (CPR), RIKEN iTHEMS, Wako, Saitama 351-0198, Japan
}

\begin{abstract}
\textbf{
{
Extending notions of phase transitions to nonequilibrium realm is a fundamental problem for statistical mechanics.
While it was discovered that critical transitions occur even for transient states before relaxation as the singularity of a dynamical version of free energy, their nature is yet to be elusive.
Here, we show that spontaneous symmetry breaking  can occur at a short-time regime and causes  universal dynamical quantum phase transitions in periodically driven unitary dynamics.
Unlike conventional phase transitions, the relevant symmetry is antiunitary: its breaking is accompanied by a many-body exceptional point of a nonunitary operator obtained by space-time duality.
Using a stroboscopic Ising model, we demonstrate the existence of
distinct phases and unconventional singularity of dynamical free energy, whose signature can be accessed through quasilocal operators.
Our results open up research for hitherto unknown phases in short-time regimes, where time serves as another pivotal parameter, with their hidden connection to nonunitary physics.
}
}

\end{abstract}

\maketitle

Phase transition~\cite{landau2013statistical,sachdev2007quantum} is one of the most fundamental collective phenomena in macroscopic systems.
%While equilibrium phase transition is characterized by singularities of free energies, r
Recent experiments on artificial quantum many-body systems motivate researchers to understand phases and their transitions in systems out of equilibrium.
Various nonequilibrium phases are proposed including e.g., many-body localized phases~\cite{Schreiber15,Smith16}, Floquet topological phases~\cite{lindner2011floquet,rechtsman2013photonic}, and discrete time crystals~\cite{Choi17,Zhang17,Moessner17}.

{
Recently, dynamical quantum phase transitions (DQPTs) particularly gather great attention as a nonequilibrium counterpart of equilibrium phase transition, which occurs for transient times of quantum relaxation~\cite{Heyl13,Heyl18}.
Defined as the singularity of the so-called dynamical free energy (especially at critical times), which is calculated from the overlap between the time-evolved and reference states, the {DQPT} has been actively studied theoretically~\cite{Karrasch13,Heyl14,Andraschko14,Canovi14,Heyl15,Sharma15,Budich16,Sharma16,Halimeh17,Zauner-Stauber17,Zunkovic18,Bhattacharyya20,Bandyopadhyay20,Halimeh20} and experimentally~\cite{Jurcevic17,Flaschner18}.
}

{
Despite extensive studies, the nature of DQPTs is yet to be elusive.
One of the important problems is what mechanism leads to DQPTs.
Several studies find that some DQPTs are associated with equilibrium/steady-state phase transition~\cite{Heyl14,Zunkovic18}.
On the other hand, DQPTs without such relations may also exist~\cite{Andraschko14,Halimeh17}, which 
 indicates that DQPTs can be caused by an unconventional mechanism unique to finite-time (high-frequency) regime of quantum relaxation.
Another open problem is universality and criticality of DQPTs.
Although typical DQPTs are accompanied by cusps of dynamical free energies~\cite{Heyl13,Heyl18}, several works report DQPTs with different types of singularities~\cite{Heyl15,Bhattacharyya20}.
However, a clear understanding of universality and criticality of DQPTs is far from complete.
}
\begin{figure}
\begin{center}
\includegraphics[width=\linewidth]{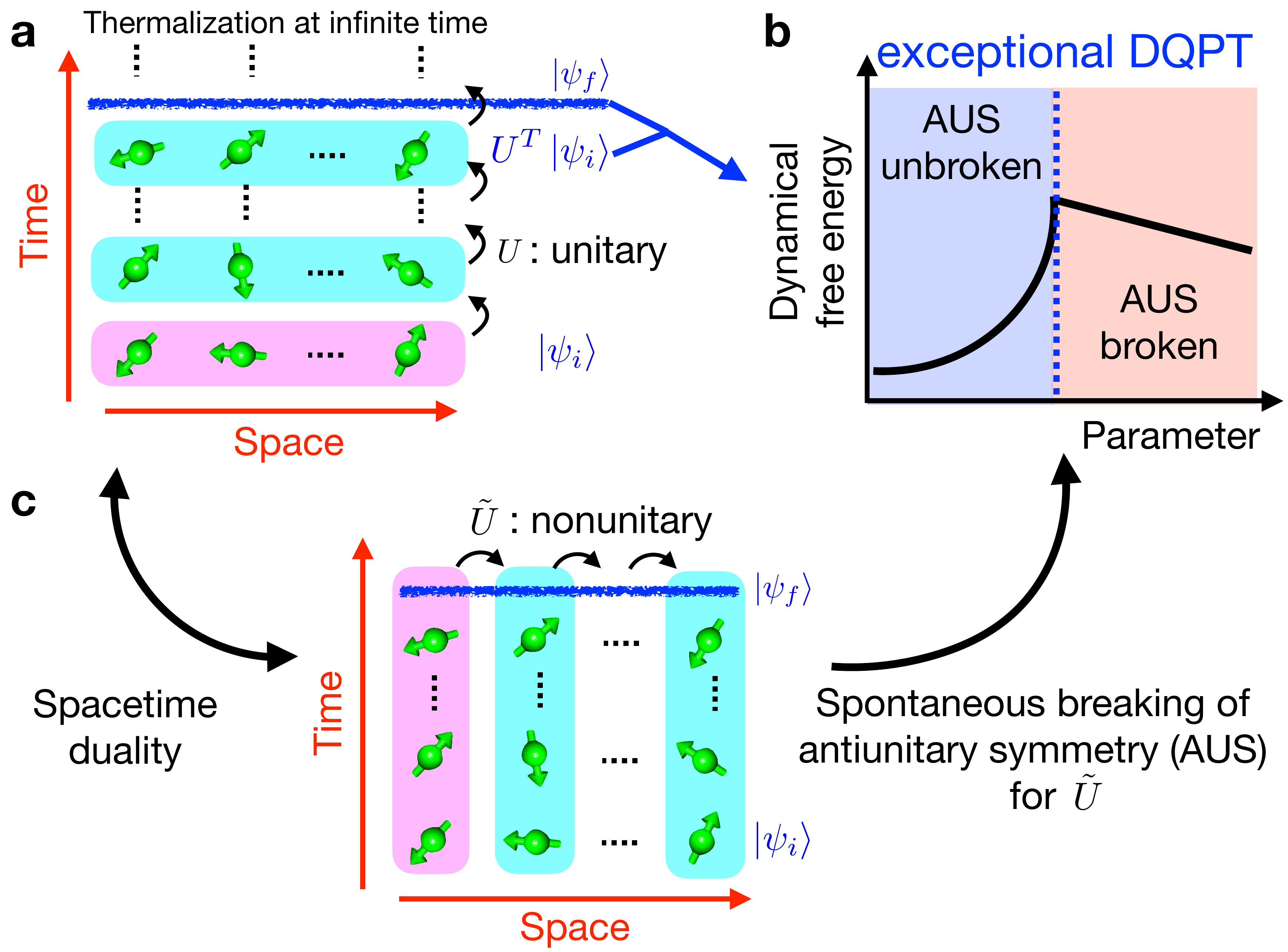}
\end{center}
\caption{
{
\textbf{Schematic of the exceptional dynamical quantum phase transition (DQPT) and its origin as the spontaneous antiunitary symmetry (AUS) breaking.}
\textbf{a,} Periodically driven unitary dynamics described by an operator $U$. While our nonintegrable system thermalizes for local observables at infinite times, we here discuss the phase transitions occurring at finite times.
\textbf{b,} Example of the exceptional DQPT.
Dynamical free energy, which is obtained from the overlap between time-evolved state $U^T\ket{\psi_i}$ and the reference state $\ket{\psi_f}$, has a divergent derivative at some critical parameter approached from the AUS-unbroken phase.
\textbf{c,} Origin of the exceptional DQPT. 
Using the spacetime duality, we find a hidden nonunitary transfer operator $\tilde{U}$ that propagates in space direction.
We uncover that the exceptional DQPTs arise when the AUS for $\tilde{U}$ is spontaneously broken.
}
}
\label{intro}
\end{figure}\\
{\quad
In this work, we find universal DQPTs in periodically driven unitary dynamics
caused by the spontaneous ``antiunitary symmetry (AUS)" breaking.
While spontaneous symmetry breaking is a fundamental mechanism for conventional phase transitions, several distinct features appear in our results.
First, the AUS breaking in our model occurs uniquely at finite times and cannot be captured by conventional equilibrium or steady-state phases.
Second, the AUS appears as a symmetry of a hidden nonunitary transfer operator, which is obtained by switching the role of space and time.
Consequently, the universality and criticality found in the unitary dynamics are characterized by those of the exceptional point, which recently gathers great attention in non-Hermitian physics~\cite{ElGanainy18,Miri19}; thus we call the transition the exceptional DQPT.
To demonstrate our discovery, we particularly use a stroboscopic chaotic Ising chain and show that the derivative of dynamical free energy defined at finite times can diverge through changes of a  parameter  (Fig.~\ref{intro}\textbf{a,b}).
Using the recently developed technique called the spacetime duality~\cite{Akila16,Bertini18,Bertini19,Bertini19EC} and determining the hidden nonunitary operator, we discuss several properties of the exceptional DQPT besides the divergence of the dynamical free energy  (Fig.~\ref{intro}\textbf{c}).
For example, instead of the long-range order associated with conventional symmetry breaking, we show that the generalized correlation function has the divergent correlation length at transition and exhibits oscillatory long-range order after antiunitary symmetry breaking.
Finally, we demonstrate that the signatures of the exceptional DQPTs are observed through quasi-local observables that are accessible by state-of-the-art experiments~\cite{Zhang17,Tan21}.
Notably, we argue that the signature of the exceptional DQPTs is easier to observe than that of the normal DQPTs because of their strong singularity.
Our results make an important step toward understanding the nature of phase transitions occurring in a short-time regime, which goes beyond conventional phase transitions since time serves as another crucial  parameter here, with their hidden connection to nonunitary physics.
}
\newline

{\textbf{\large{Results}}}

\textbf{Stroboscopic Ising chains and dynamical free energy.}
To demonstrate our finding, we introduce a one-dimensional quantum stroboscopic spin model{~\cite{Prosen02,Akila16,Bertini18}} composed of Ising interaction and subsequent global rotation.
{This model is a prototypical model for quantum chaotic dynamics and can be realized in experiments of e.g., trapped ions~\cite{Zhang17}.}
Its unitary time evolution for a single step can be written as
\aln{
U=e^{-i\sum_{j=1}^L b\sigma_j^x}e^{-i\sum_{j=1}^LJ\sigma_j^z\sigma_{j+1}^z-i\sum_{j=1}^Lh\sigma_j^z},
}
where we impose a periodic boundary condition.

Let us consider a time-evolved state $U^T\ket{\psi_i}$ after $T$ steps from an initial state $\ket{\psi_i}$.
To characterize this nonequilibrium state, we focus on  the overlap with another state $\ket{\psi_f}$, i.e.,
$\braket{\psi_f|U^T|\psi_i}$.
The logarithm of the absolute value of this overlap per system size, $F_{L,T}$, is dubbed as the dynamical free energy density~\cite{Heyl18}.
{We here consider three types of dynamical free energy density.
The first one is to take $\ket{\psi_i}=\ket{\psi_f}=\ket{\psi}$ and average the overlap over  $\ket{\psi}$ randomly taken from the unitary Haar measure before taking the absolute value and the logarithm.}
Then, the (modified) dynamical free energy density reads 
\aln{
F_{L,T}^\mr{Tr}(b,J,h)=-\frac{1}{L}\log|\Tr[U^T]|+\log 2.
}
{We note that $F_{L,T}^\mr{Tr}$ is the logarithm of the two-point spectral measure through $|\Tr[U^T]|^2=\sum_{a,b}e^{iT(z_a-z_b)}$, where $e^{iz_a}$ are the eigenvalues for $U$.
Since the appearance of trace simplifies the discussion, we mainly use this quantity to show our results.}

{The second one is to take $\ket{\psi_i}= \bigotimes_{j=1}^L\ket{\uparrow_j}$ and $\ket{\psi_f}= \bigotimes_{j=1}^L\ket{\downarrow_j}$, where $\ket{\uparrow_j}$/$\ket{\downarrow_j}$ is the eigenstate of $\sigma_j^z$ with an eigenvalue $+1$/$-1$.
In this case, we have
\aln{\label{downup}
F^{\downarrow\uparrow}_{L,T}(b,J,h)=-\frac{1}{L}\log|\braket{\downarrow\cdots \downarrow|U^T|\uparrow\cdots \uparrow}|.
}
The third one is to take $\ket{\psi_i}=\ket{\psi_f}= \bigotimes_{j=1}^L\ket{\uparrow_j}$, leading to 
$F^{\uparrow\uparrow}_{L,T}(b,J,h)=-\frac{1}{L}\log|\braket{\uparrow\cdots \uparrow|U^T|\uparrow\cdots \uparrow}|$.
}

The derivative of $F_{L,T}$ gives the (imaginary part of) so-called generalized expectation values. For example, we have 
\aln{\label{gexp}
-\frac{1}{T}\fracd{F^\mr{Tr}_{L,T}}{b}=\mr{Im}\lrl{\Tr\lrs{\frac{1}{L}\sum_j\sigma_j^x\tilde{\rho}}}:=\mr{Im}\lrl{\braket{\sigma^x_1}_\mr{gexp}},
}
where $\tilde{\rho}=U^T/\Tr[U^T]$ 
%is not a quantum state since it lacks positivity in general, 
and we have used translation invariance.
Importantly, the dynamical free energy density and the generalized expectation values can be in principle measured with an interferometric experiment~\cite{Canovi14,Heyl18}.

We seek for singularities of $F_{\infty,T}$ when some continuous parameter is varied.
In Ref.~\cite{Heyl13}, $F_{\infty,T}$ exhibits singularity at critical times for continuous-time models.
Since $T$ is discrete in our model, instead of changing $T$, we consider continuously changing other parameters (such as $b$) for fixed $T$.

\begin{figure}
\begin{center}
\includegraphics[width=\linewidth]{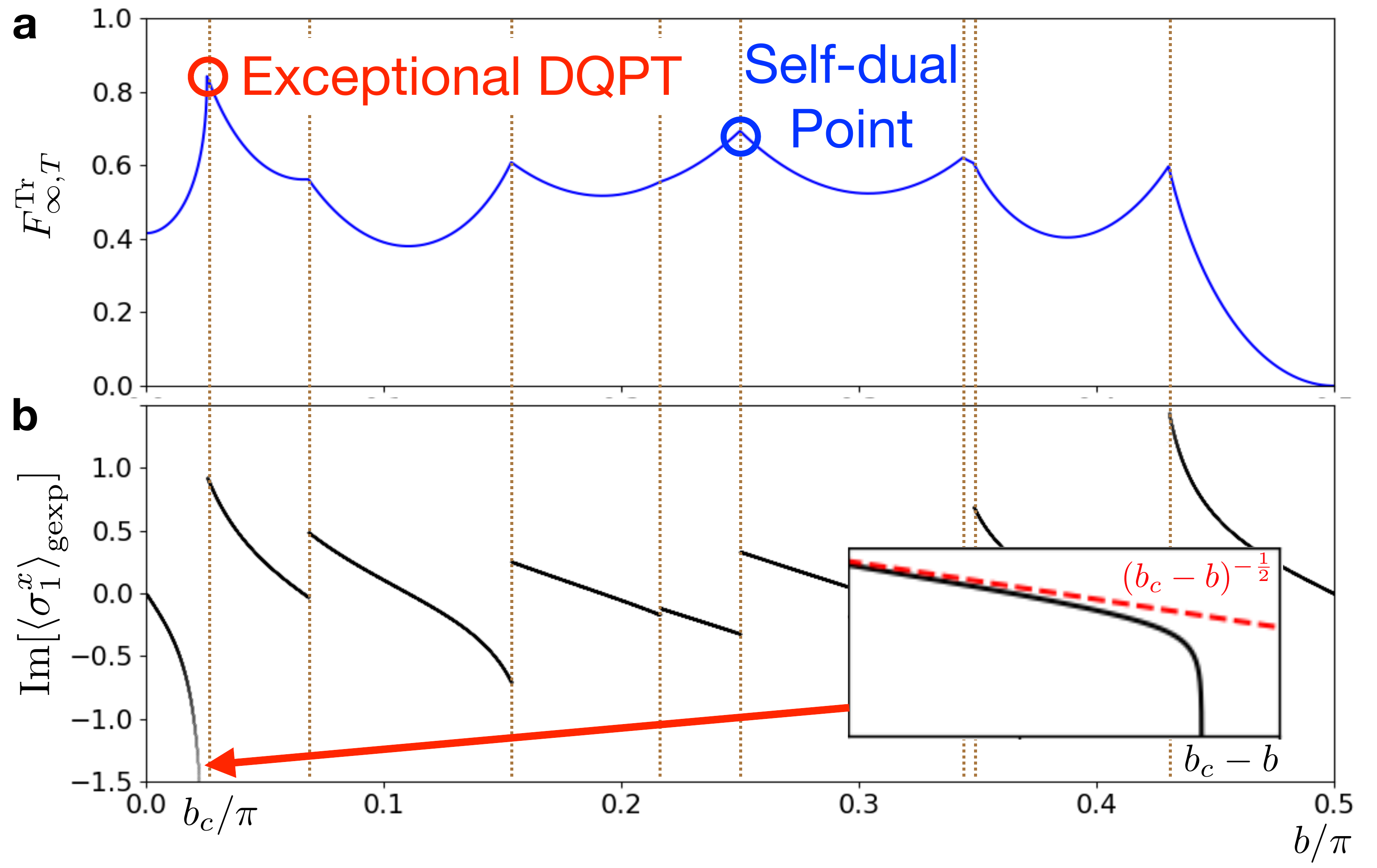}
\end{center}
\caption{
\textbf{Occurrence of {DQPTs} caused by the variation of the rotation angle $b$.}
\textbf{a,} Dynamical free energy density $F^\mr{Tr}_{\infty,T}$, whose singularities indicate {DQPTs}.
Among {DQPT}s, we have the exceptional {DQPT} (red circle), which shows divergent derivative, and the {DQPT} crossing the self-dual point (blue circle).
\textbf{b,} Imaginary part of the generalized expectation value given in Eq.~\eqref{gexp}, which is proportional to the $b$-derivative of $F^\mr{Tr}_{\infty,T}$.
While it exhibits a jump for typical {DQPT}s, it diverges at the exceptional {DQPT}.
(inset) The divergence obeys $(b_c-b)^{-1/2}$ (red dashed line).
We use $J=-\pi/4$ and $h=3.0$, and $T=6$.
}
\label{edpqt}
\end{figure}

\textbf{Dynamical phases and their transitions.}
As a prime example that highlights our discovery, we show  in Fig.~\ref{edpqt}  the (real-part of) dynamical free energy density $F^\mr{Tr}_{\infty,T(=6)}$ and $\mr{Im}[\braket{\sigma_1^x}_\mr{gexp}]$ as a function of the rotation angle $b$ for $J=-\pi/4$ and $h=3.0$ (see  Supplementary Note 1 for the data with other parameters and initial/final states).
This is calculated from the eigenvalue with the largest modulus of the space-time dual operator, as detailed later.
We find different singular behaviors for $F_{\infty,T}^\mr{Tr}$, signaling distinct {DQPTs} at critical parameters.
{
Many cusps of $F^{\mr{Tr}}_{\infty,T}$ with varying $b$ are are analogous to (continuous time) {DQPT}s studied previously, where $\mr{Im}[\braket{\sigma_1^x}_\mr{gexp}]$  exhibits a finite jump.}

Notably, we find a distinct singularity at $b=b_c\simeq 0.0257$ for $T=6$, where the derivative diverges as $\fracd{F^\mr{Tr}_{\infty,T}}{b}\propto \mr{Im}[\braket{\sigma_1^x}_\mr{gexp}]\sim |b_c-b|^{-1/2}$ for $b\lesssim b_c$. 
Such a strong singularity is prohibited for equilibrium free energy density since the thermal expectation value of a local observable cannot diverge.
We call this transition an {exceptional {DQPT}}, as it turns out to originate from the occurrence of an exceptional point of a nonunitary operator that is dual to $U$.
As shown below, an exceptional {DQPT} can occur for $F^{\{\mr{Tr}/\downarrow\uparrow\}}_{\infty,T}$ with $J=\frac{\pi}{4}+\frac{n\pi}{2}\:(n\in\mbb{Z})$ and even/odd $T$ and is robust under certain weak perturbation (such as $h$), which is deeply related to the hidden symmetry of our setup.
{We note that the value of $b_c$ itself depends on the parameters, such as $T$.
We also note that, while the divergence of the derivative of dynamical free energy was recently found in Ref.~\cite{Bhattacharyya20} for an integrable system, the connection to the underlying symmetry was not discussed.}

{
The exceptional {DQPT} occurs at a different point from the self-dual points, which are $J=\frac{\pi}{4}+\frac{n\pi}{2}$ and $b=\frac{\pi}{4}+\frac{m\pi}{2}\:(n,m\in\mbb{Z})$ and known in the context of quantum many-body chaos~\cite{Bertini18,Bertini19}.
As discussed in  Supplementary Note 5, we find that crossing self-dual points entails {DQPT} universally for $F^{\mr{Tr}/\uparrow\uparrow/\downarrow\uparrow}_{\infty,T}$ with any $T$ and $h$, whose criticality is analogous to that for the conventional {DQPT} (see Fig.~\ref{edpqt}).}

{
We stress that {DQPT}s in our model do not appear as  infinite-time averages of expectation values of local observables (see  Supplementary Note 4), in contrast with the observation in Ref.~\cite{Zunkovic18}.
Indeed, our {DQPT}s occur at nonintegrable points, where the infinite-time averages of expectation values trivially thermalize because of the Floquet eigenstate thermalization hypothesis~\cite{Kim14}.
This means that our {DQPT}s are unique to finite-time regimes, in which time serves as an important parameter in stark contrast with conventional phase transitions.
}

\textbf{Spacetime duality and hidden symmetries.}
To understand the above behaviors, we employ the space-time duality~\cite{Akila16} of our Floquet operator.
This is an exact method to switch the role of time and space and rewrite $U^T$ with $L$ product of  a space-time-dual transfer matrix $\tilde{U}$, which involves $T$ spins.
{Using this method, we can  rewrite the dynamical free energy as
\aln{
F_{L,T}=-\frac{1}{L}\log|\Tr[\tilde{U}^L]|,
}
 where the nonunitary operator $\tilde{U}$ depends on the type of $F_{L,T}$.}
 For example, we have
\aln{
\tilde{U}_\mr{Tr}=Ce^{-i\sum_{\tau=1}^T \tilde{b}\sigma_\tau^x}e^{-i\sum_{\tau=1}^T\tilde{J}\sigma_\tau^z\sigma_{\tau+1}^z-i\sum_{\tau=1}^Th\sigma_\tau^z}
}
with the periodic boundary condition
for $F^\mr{Tr}_{L,T}$~\cite{Akila16,Bertini18,Bertini19} (see  Supplementary Note 2 for the proof and the similar construction for ${\tilde{U}}_{\uparrow\uparrow/\downarrow\uparrow}$, which corresponds to $F^{\uparrow\uparrow/\downarrow\uparrow}_{L,T}$).
%, and
%\aln{
%\tilde{U}_{\uparrow\uparrow/\downarrow\uparrow}=C'e^{-i\sum_{\tau=1}^{T-1} \tilde{b}\sigma_\tau^x}e^{-i\sum_{\tau=1}^{T-2}\tilde{J}\sigma_\tau^z\sigma_{\tau+1}^z-i\sum_{\tau=1}^{T-1}h\sigma_\tau^z-i\tilde{J}(\sigma_1^z+I\sigma_{T-1}^z)}
%}
%with the open boundary condition for $F^{\uparrow\uparrow/\downarrow\uparrow}_{L,T}$.
Here, $\tilde{b}=-\pi/4-i\log(\tan J)/2$, $\tilde{J}=-\pi/4-i\log(\tan b)/2$ and $C=(\sin 2b/\sin 2\tilde{b})^{T/2}/2$.
%$C'=(\sin 2b/2i)^{1/2}(\sin 2b/\sin 2\tilde{b})^{(T-1)/2}e^{-i(h+J)}$
%, and $I=1\:(-1)$ for $F^{\uparrow\uparrow}_{L,T}\:(F^{\downarrow\uparrow}_{L,T})$.

\begin{figure}
\begin{center}
\includegraphics[width=\linewidth]{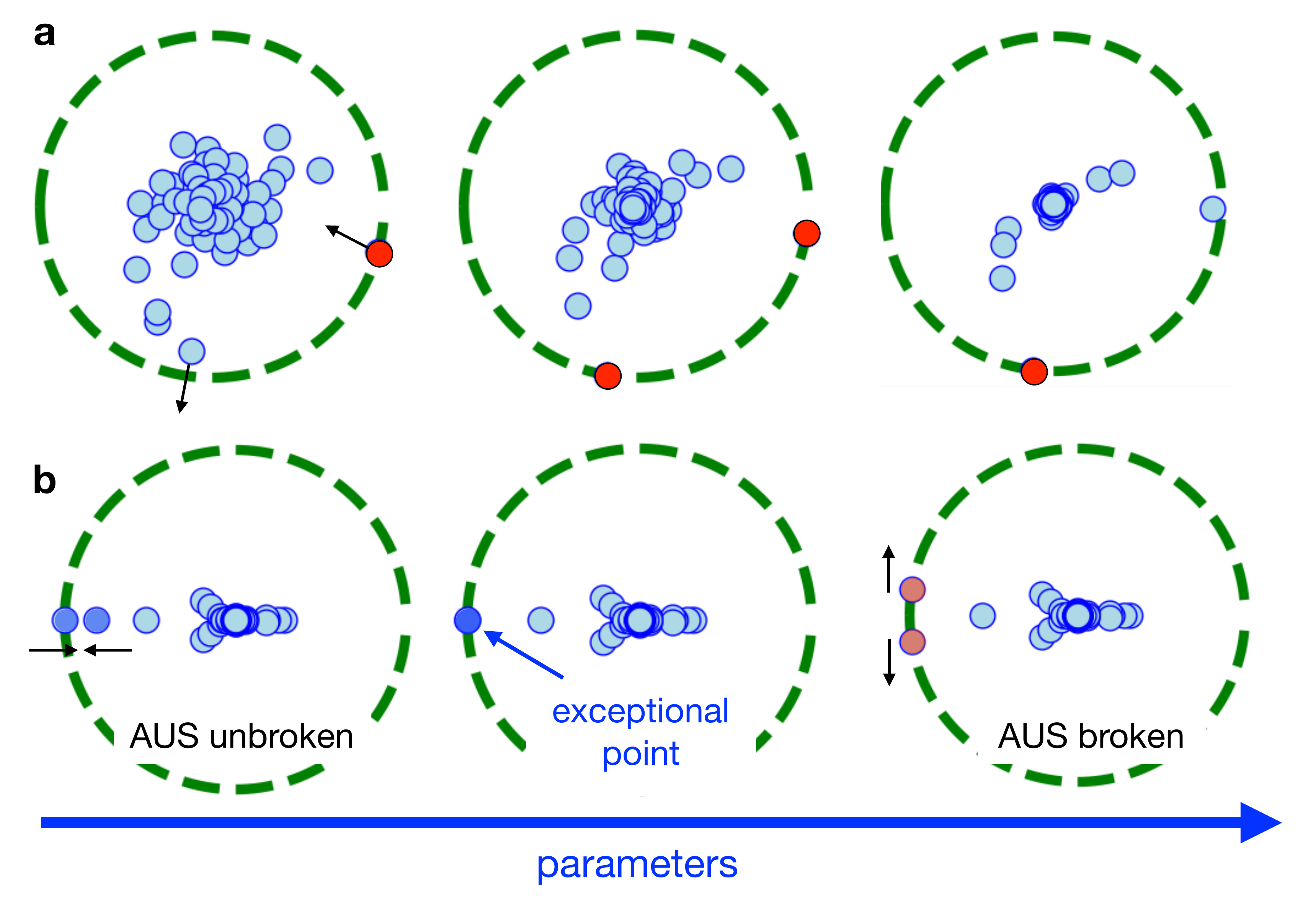}
\end{center}
\caption{
\textbf{Schematic of eigenvalue dynamics of the spacetime-dual operator $\tilde{U}$ and its relation to the {DQPT}.}
\textbf{a,} Typical eigenvalue dynamics  (small circles) near {DQPT}.
Green dashed circles have the radius that corresponds to the eigenvalue(s) with the largest modulus.
The eigenvalue with the largest modulus (red circles) switches at the critical point, at which two eigenvalues have the same modulus.
{
\textbf{b,} Eigenvalue dynamics through the exceptional {DQPT}.
Eigenvalues with the largest and the second-largest modulus lie on the same radial direction protected by antiunitary symmetry (AUS) of $\tilde{U}$ when  AUS is unbroken.
When the parameter changes, the eigenvalues coincide at the critical parameter and show spectral singularity as an exceptional point.
They then form a complex-conjugate pair (i.e., AUS breaking) and the modulus of two eigenvalues becomes equivalent.}
%At the self-dual point, all of $2^T$ numbers of eigenvalues have the same modulus owing to the unitarity of $\tilde{U}$.
}
\label{eigenvalue}
\end{figure}

Let $\lambda_\mr{M,\alpha}=|\lambda_\mr{M}|e^{i\theta_\alpha}$  be eigenvalues of $\tilde{U}$ whose modulus gives the largest one among all eigenvalues.
Here, $\alpha\:(=1,\cdots, n_\mr{deg})$ is the label of the degeneracy, where $n_\mr{deg}$ is the number of eigenvalues giving the maximum modulus.
For large $L$, $F_{L,T}$  is dominated by these largest eigenvalues, i.e.,
\aln{\label{eq:fltdual}
F_{L,T}\simeq -\log |\lambda_\mr{M}|-\frac{1}{L}\log\lrv{\sum_\alpha e^{i\theta_\alpha L}}.
}
In the thermodynamic limit, the second term vanishes.

{Similar to the discussion noted in Ref.~\cite{Andraschko14},}  {DQPTs} occur when the eigenstate that gives the largest eigenvalue switches.
For typical cases, conventional {DQPTs} occur when maximum of two eigenvalues with different $\theta_\alpha$ switches accidentally, where $n_\mr{deg}=1$ for each phase and $n_\mr{deg}=2$  at transition  (Fig.~\ref{eigenvalue}\textbf{a}).

In contrast, hitherto unknown dynamical phases and transitions can appear when $\tilde{U}$ possesses AUS~\cite{Gong18T,Kawabata18T, Kawabata19S}.
{In nonunitary physics, the operator $\tilde{U}$ is said to have the  AUS when some unitary operator $V$ and $\phi\in\mbb{R}$ exist and $V\tilde{U}^*V^\dag=e^{i\phi}\tilde{U}$ is satisfied (see Table~I).
As detailed in the Methods section,  nonunitary operator $\tilde{U}$ is called Class A if $\tilde{U}$ does not have the AUS, Class AI when the AUS exists and the corresponding  $V$ satisfies $VV^*=\mbb{I}$, and Class AII when the AUS exists and the corresponding  $V$ satisfies $VV^*=\mbb{I}$.}
{
A particularly important class is Class AI, where the spectral transition unique to nonunitarity, i.e., spontaneous AUS breaking, occurs with the change of parameters.
In this case, the eigenstates do and do not respect the AUS for each phase separated at the critical point, which is called the exceptional point.
Through the transition, two eigenvalues are attracted, degenerated (at the exceptional point), and repelled in a singular manner (see Fig.~\ref{eigenvalue}\textbf{b}).}

We find that some of our Floquet operators $U$ can have hidden AUS of $\tilde{U}$ for $J=\frac{\pi}{4}+\frac{n\pi}{2}\:(n\in\mbb{Z})$ (see Table~II and Methods section).
Particularly, $\tilde{U}_\mr{Tr}$ belongs to Class AI for even $T$ (and AII for odd $T$), and $\tilde{U}_{\downarrow\uparrow}$ belongs to Class AI for odd $T$ (and AII for even $T$) as long as $J=\frac{\pi}{4}+\frac{n\pi}{2}\:(n\in\mbb{Z})$.
In contrast, $\tilde{U}_{\uparrow\uparrow}$ does not have AUS and belongs to Class A in general.

\begin{table}
{
 \caption{Summary of antiunitary symmetry (AUS) classes. Only Class AI can exhibit the AUS-breaking transition.}}
 \centering
  \begin{tabular}{|c|c|c|}
   \hline
AUS class&  $V\tilde{U}^*V^\dag=e^{i\phi}\tilde{U}$?& AUS-breaking transition\\
   \hline
   Class A& No & No\\
   \hline
Class AI& $VV^*=\mbb{I}$ &Yes\\
   \hline
Class AII& $VV^*=-\mbb{I}$&No \\
   \hline
  \end{tabular} \label{AUSclass}
\end{table}

\begin{table}{
 \caption{Summary of antiunitary symmetry classes for each operator $\tilde{U}$ with $J=\frac{\pi}{4}+\frac{n\pi}{2}\:(n\in\mbb{Z})$. The AUS-breaking transition can occur for Class AI, which corresponds to $\tilde{U}_\mr{Tr}$ with even $T$ and $\tilde{U}_\mr{\downarrow\uparrow}$ with odd $T$.}}
 \label{Uop}
 \centering
  \begin{tabular}{|c|c|c|}
   \hline
$\tilde{U}$ with $J=\frac{\pi}{4}+\frac{n\pi}{2}$ &  even $T$ & odd $T$\\
   \hline
$\tilde{U}_\mr{Tr}$& Class AI & Class AII\\
   \hline
$\tilde{U}_\mr{\downarrow\uparrow}$& Class AII & Class AI\\
   \hline
$\tilde{U}_\mr{\uparrow\uparrow}$& Class A&Class A \\
   \hline
  \end{tabular}
\end{table}

The above symmetries clearly explain the origin of the exceptional {DQPT}:
as shown in Fig.~\ref{eigenvalue}\textbf{b}, this transition occurs when eigenvalues with the largest and the second-largest modulus collide under Class AI AUS, i.e., at the many-body exceptional point~\cite{Ashida17,Hamazaki19N,Luitz19} for $\lambda_\mr{M}$.
It is known that this (second-order) exceptional point entails a universal spectral singularity, where the gap between two eigenvalues behave like $|b-b_c|^{1/2}$.
%This leads to the previously-mentioned notable divergence of the generalized expectation value $\sim(b_c-b)^{-1/2}$ for $b<b_c$~\cite{derivative-EDQPT}, where $-1/2$ is also known to be a universal critical exponent.
This leads to the previously-mentioned notable divergence of the generalized expectation value $\sim(b_c-b)^{-1/2}$ for $b<b_c$, where $-1/2$ is also known to be a universal critical exponent.
%In Fig.~\ref{exceptional}, we show  the gap between the two eigenvalues and the overlap of these eigenstates, both of which ensure the existence of the exceptional point.

\begin{figure}
\begin{center}
\includegraphics[width=\linewidth]{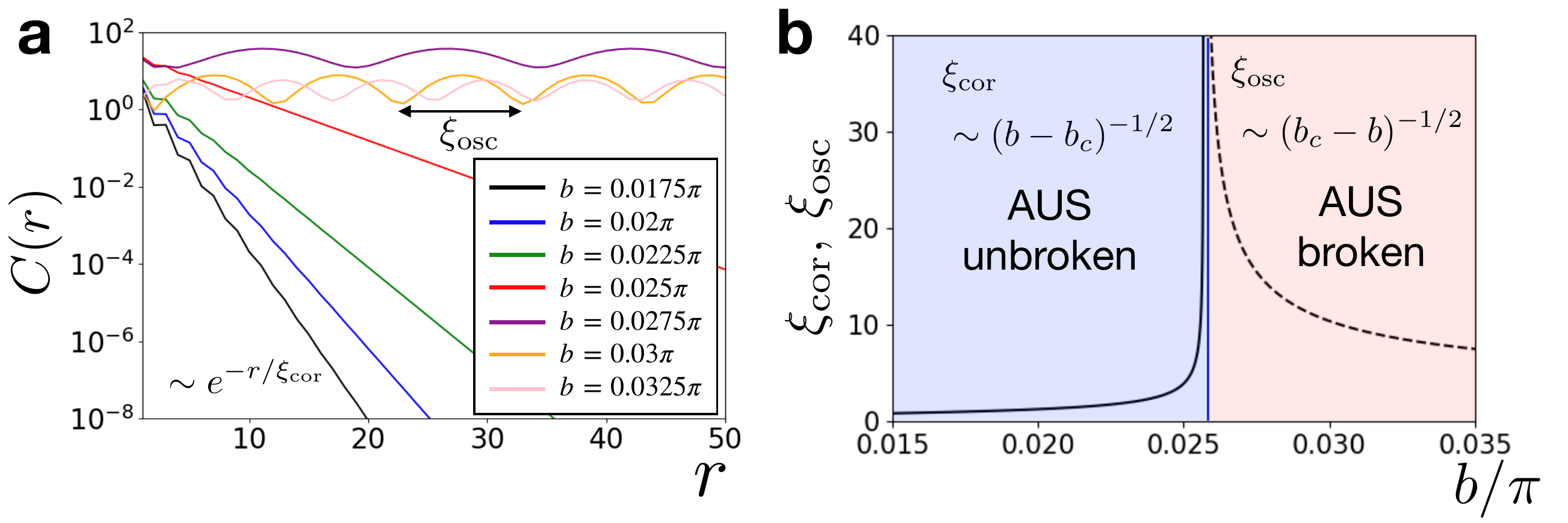}
\end{center}
\caption{
\textbf{Generalized correlation function  and correlation/oscillation lengths $\xi_\mr{cor},\xi_\mr{osc}$ corresponding to $F^\mr{Tr}_{\mr{\infty},{T}}$.}
\textbf{a,} Generalized correlation function $C(r)$ for different values of $b$ for $L=500$.
In the AUS-unbroken phase ($b<b_c\simeq 0.0257\pi$), the correlation decays exponentially.
In the AUS-broken phase ($b>b_c$), the correlation exhibits oscillatory long-range order.
\textbf{b,} Divergence of $\xi_\mr{cor}$ (solid line) and $\xi_\mr{osc}$ (dotted line) in the thermodynamic limit. Approaching the exceptional {DQPT}, they both behave as $\sim |b-b_c|^{-1/2}$.
We use $J=-\pi/4$ and $h=3.0$, and $T=6$.
}
\label{correlation}
\end{figure}

For $F^\mr{Tr}_{\mr{\infty},{T=6}}$, the phases with $b<b_c\simeq 0.0257\pi$ and $b>b_c$ correspond to hidden AUS-unbroken and AUS-broken phases, respectively.
This is highlighted by the generalized correlation function, $C(r)=|\braket{\sigma_1^z\sigma_{r+1}^z}_\mr{gexp}-\braket{\sigma_1^z}_\mr{gexp}\braket{\sigma_{r+1}^z}_\mr{gexp}|$ (see Fig.~\ref{correlation} and the Methods section).
While $C(r)$ decays exponentially as $\sim e^{-r/\xi_\mr{cor}}$ in the AUS-unbroken phase, the correlation length diverges as $\xi_\mr{cor}\sim (b_c-b)^{-1/2}$ as it approaches the exceptional {DQPT} point.
At AUS-broken phases, $\xi_\mr{cor}$ diverges and long-range order appears.
Notably, we find that $C(r)$ oscillates with the oscillation length $\xi_\mr{osc}$, which also diverges near the exceptional {DQPT} $\xi_\mr{osc}\sim (b-b_c)^{-1/2}$.
 %The damping-oscillation transition is similar to the non-Hermitian dynamics with parity-time-symmetry breaking, but it appears as the long-range order in space because the symmetry breaking occurs after the spacetime duality.
We remark that the qualitative signature of the transition can be captured by the existence of the long-range order even for relatively small systems, which are relevant for experiments (see  Supplementary Note 6).

{
Here, we comment on the relation with the seminal work by Lee, Yang~\cite{Lee52,Yang52} and Fisher~\cite{Fisher78}, who investigated thermodynamic phase transitions by non-Hermitian operators.
While our motivation is to investigate {DQPT}s occurring at finite times, which is different from their motivation, there exists some mathematical analogy.
In fact, the exceptional {DQPT} can be regarded as the realization of the edge singularity of the partition-function zeros at physical (i.e., real) parameters, as discussed in the Methods section.
}

Hidden Class AI AUS also enables us to discuss conditions for having exceptional {DQPTs}.
In our prototypical stroboscopic Ising model, we can observe the exceptional {DQPT} by considering $F^\mr{Tr}_{\infty,T}$ with even $T$ and  $F^{\downarrow\uparrow}_{\infty,T}$ with odd $T$ under the condition $J=\pi/4+n\pi/2\:(n\in\mbb{Z})$ {(see  Supplementary Note 3 for the example of $F^{\downarrow\uparrow}_{L,T}$).}
Note that this transition is robust even if the value of $h$ is slightly perturbed since the transition is protected by AUS.
{We also stress that $J$ cannot be generic in our anlysis: $J=\pi/4+n\pi/2\:(n\in\mbb{Z})$ is important for the exceptional {DQPT} because it ensures the antiunitary symmetry for $\tilde{U}$.
Investigation of the exceptional {DQPT} for other values of $J$ is a future problem.}
%While the small perturbation on $J$ eliminates the exact transition, the signal of divergence at the exact transition appears as a peak for the generalized observable.

\textbf{Signature through quasi-local observables.}
{
Next, we show that the signature of our {DQPT}s is accessible through the expectation values of quasi-local observables, which are more experimentally friendly than the overlap itself (in other words, the {DQPT} affects the behavior of the expectation values of the quasi-local observables).
We also demonstrate that the exceptional {DQPT} is easier to measure with finite-size scaling analysis than the conventional {DQPT}, thanks to its strong singularity.
We here explain this fact by focusing on $F^{\downarrow\uparrow}_{L,T}$ in Eq.~\eqref{downup}, instead of $F^{\Tr}_{L,T}$, since its operational meaning in  experimental situations is more direct.
We note that $F^{\downarrow\uparrow}_{\infty,T}$ shows the exceptional DQPT for $b=b_c\simeq 0.446\pi$ with $h=1.3$, $T=5$ and $J=-\pi/4$, where the AUS is broken for $b<b_c$ and unbroken for $b>b_c$ (this is opposite to the case for $F^\mr{Tr}_{\infty,T}$).}

{
To see our argument, we introduce the following quantity
\aln{\label{Fmod2}
F^{\downarrow\uparrow(l)}_{L,T}=-\frac{1}{2l}\log \braket{\psi_i|P_f^{(l)}(T)|\psi_i},
}
where
 $
P_f^{(l)}=\bigotimes_{i=1}^l\ket{\downarrow}_i\bra{\downarrow}_i
$
and $P_f^{(l)}(T)=U^{-T}P_f^{(l)}U^{T}$ is the Heisenberg representation.
While $P_f^{(l=L)}=\bigotimes_{i=1}^L\ket{\downarrow}_i\bra{\downarrow}_i=\ket{\psi_f}\bra{\psi_f}$ and Eq.~\eqref{Fmod2} reduces to $F^{\downarrow\uparrow}_{L,T}$ for $l=L$,
$P_f^{(l)}$ becomes quasi-local when $l=\mr{O}(L^0)\ll L$~\cite{Halimeh20, Bandyopadhyay20}.
For the latter case,  Eq.~\eqref{Fmod2} is represented by the standard expectation value of the quasi-local observable, which describes the presence of consecutive spin-down domain at size $l$, at time $T$.
Note that such spin domains have been measured in ion experiments using single-site imaging~\cite{Zhang17,Tan21}.}

\begin{figure}
\begin{center}
\includegraphics[width=\linewidth]{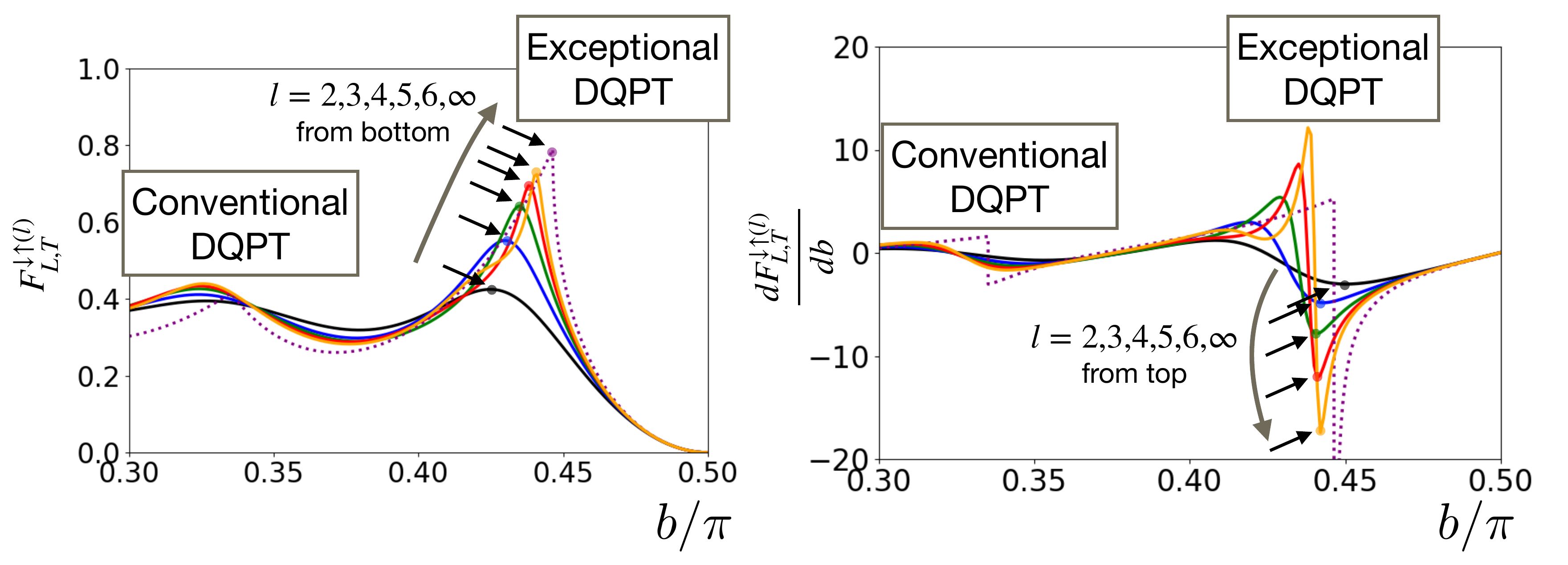}
\end{center}
\caption{
{\textbf{The signature of the exceptional {DQPT} using $F^{\downarrow\uparrow(l)}_{L,T}$ and its derivative, which are described by the expectation value of quasi-local observables.}
We find that the peaks (indicated by the dots and arrows) develop even for small $l$ around the exceptional {DQPT} ($b\simeq 0.44\pi$).
Particularly, the peaks for the derivative become rapidly sharper as increasing $l$, reflecting the divergence for $l=L\ra\infty$.
This is in contrast with the conventional {DQPT} ($b\simeq 0.33\pi$), where we cannot find sharp peaks for $l\leq 6$.
We use $L=100$ for ($l=2,3,4,5,6$) and $L=\infty$ for $l=\infty$, $T=5$, $J=-0.25\pi$ and $h=1.3$.}
}
\label{fig3}
\end{figure}

{
We argue that the signature of the exceptional {DQPT} can be captured by $F^{\downarrow\uparrow(l)}_{L,T}$ and its derivative even for relatively small $l$, which is more experimentally friendly than the dynamical free energy density itself.
Figure~\ref{fig3} shows the $b$-dependence of  $F^{\downarrow\uparrow(l)}_{L,T}$ and $\partial F^{\downarrow\uparrow(l)}_{L,T}/\partial b$  for different $l\:(=2,3,4,5,6, \infty)$.
We find that the peak develops even for small $l$ around the exceptional {DQPT} ($b\simeq 0.44\pi$).
Particularly, the peaks for the derivative become rapidly sharper as increasing $l$, reflecting the divergence for $l=L\ra\infty$.
Our results physically mean that, in this setting, large spin-down domains are rapidly suppressed toward the exceptional {DQPT} critical point.}

{
We also note that the sharp peaks indicate the experimental advantage of considering the exceptional {DQPT} compared with the conventional {DQPT}.
Indeed, as shown in Fig.~\ref{fig3}, we cannot find sharp peaks for $l\leq 6$ for the conventional {DQPT} ($b\simeq 0.33\pi$).
This indicates that the exceptional {DQPT} is easier to detect even with small $l$ than the conventional {DQPT} because of its unique singularity, which is another advantage for our analysis.}
\newline

{\textbf{\large{Discussions}}}

{
Although we have demonstrated the singularity of the dynamical free energy and the oscillatory long-range order for the spontaneous antiunitary symmetry breaking, one may wonder whether we can define an order parameter that is nonzero only for the symmetry-breaking phase.
As detailed in   Supplementary Note 7, we show that an order parameter can be explicitly constructed using different-time generalized observables.
This indicates that antiunitary symmetry breaking physically means the symmetry breaking of the dynamical interference patterns, which cannot be diagnosed by the usual single-time expectation values.
}

 The exceptional {DQPT} appears in other situations as well as the above situation.
When we change $h$ instead of $b$,  AUS of $\tilde{U}_{\mr{Tr}/\downarrow\uparrow}$ is preserved and the exceptional {DQPT} appears for even/odd $T$, meaning that $\braket{\sigma^z_i}_\mr{gexp}$ diverges.
We also stress that the exceptional {DQPT} is not restricted to the stroboscopic Ising model but occurs for a broader class of Floquet systems, as shown in   Supplementary Note 8.

To conclude,
{
we have shown that the spontaneous antiunitary symmetry breaking leads to the unconventional universal DQPT,  i.e., the exceptional DQPT, uniquely at finite times in Floquet quantum many-body systems.
The appearance of finite-time phase transitions related to nonunitary physics can be understood from the spacetime duality.
We have also demonstrated that the signatures of the exceptional DQPTs are observed through quasi-local observables that are accessible by state-of-the-art experiments~\cite{Zhang17,Tan21}.
Notably, the signature of the exceptional DQPTs is easier to observe than that of the normal DQPTs because of their strong singularity.
}

{
Our result paves the way to study completely unknown phases in short-time regimes, where time is regarded as a crucial parameter.
As demonstrated in this work,} our method via spacetime duality is useful for investigating unconventional finite-time phase transitions for quantum many-body unitary dynamics through the scope of nonunitary many-body physics.
One of the promising directions is to classify such dynamical phases by the symmetries of the spacetime-dual operator {in light of non-Hermitian symmetries}, which are completely classified only recently~\cite{Gong18T,Kawabata19S}.
\newline

{\textbf{\large{Methods}}}

\textbf{Antiunitary symmetry of $\tilde{U}$.}
Let us assume that {a nonunitary operator} $\tilde{U}$ satisfies $V\tilde{U}^*V^\dag=e^{i\phi}\tilde{U}$ for some unitary operator $V$ and $\phi\in\mbb{R}$.
%With the rotation of the complex plane, this symmetry becomes essentially equivalent~\cite{Kawabata18T} to the case for $\phi=0$ for nonunitary operators, and thus we call the above symmetry as AUS.
According to the recent classification of non-Hermitian systems~\cite{Kawabata19S}, $\tilde{U}$ is called Class A without AUS, Class AI when $V$ with $VV^*=\mbb{I}$ exists, and Class AII when $V$ with $VV^*=-\mbb{I}$ exists.
If we consider $\phi=0$ without loss of generality, the eigenvalues of $\tilde{U}$ in Class AI are either real or form complex conjugate pairs.
Furthermore, at certain parameters, two real eigenvalues collide and form a complex conjugate pair, which can be called spontaneous AUS-breaking transition.
In fact, while eigenstates $\ket{\phi}$ are symmetric under AUS in the AUS-unbroken phase, i.e., $V\ket{\phi}^*=\ket{\phi}$, $V\ket{\phi}^*$ and  $\ket{\phi}$ are different int the AUS-broken phase.
At the transition point, known as the exceptional point, two eigenstates become equivalent, which offers a unique feature for nonnormal matrices.
In Class AII, on the other hand, eigenvalues generically form complex conjugate pairs  and  are not real in the presence of the level repulsion~\cite{Hamazaki20U}.

Our Floquet operators $U$ can have such hidden antiunitary symmetries of $\tilde{U}$ for $J=\frac{\pi}{4}+\frac{n\pi}{2}\:(n\in\mbb{Z})$: indeed, we find 
$
V=\prod_{\tau=1}^Te^{i\frac{\pi}{2}\sigma_\tau^y}
$
for $\tilde{U}_\mr{Tr}$ and
$
V=\mc{P}\prod_{\tau=1}^{T-1}e^{i\frac{\pi}{2}\sigma_\tau^y}
$
for $\tilde{U}_{\downarrow\uparrow}$, where $\mc{P}$ is the parity operator exchanging $\tau$ and $T-\tau$ (see  Supplementary Note 3 for the detailed calculation).
Since $VV^*$ takes either $+\mbb{I}$ or $-\mbb{I}$ depending on $T$, we find that $\tilde{U}_\mr{Tr}$ belongs to Class AI for even $T$ and AII for odd $T$, and that $\tilde{U}_{\downarrow\uparrow}$ belongs to Class AI for odd $T$ and AII for even $T$ as long as $J=\frac{\pi}{4}+\frac{n\pi}{2}\:(n\in\mbb{Z})$.
On the other hand, $\tilde{U}_{\uparrow\uparrow}$ does not have AUS and belongs to Class A in general.
\newline

\textbf{Generalized correlation function.}
To calculate the generalized correlation function, we first note the dual representation {
\aln{
C(r)=\lrv{\frac{\Tr[\tilde{U}^{L-r}\sigma^z_{\tau=1}\tilde{U}^r\sigma^z]}{\Tr[\tilde{U}^L]}-\lrs{\frac{\Tr[\tilde{U}^L\sigma^z_{\tau=1}]}{\Tr[\tilde{U}^L]}}^2}.
}
Here, we choose the time point $\tau$  for the dual spin $\sigma^z_{\tau}$ as $\tau=1$.
Inserting the eigenstate decomposition of $\tilde{U}=\sum_\alpha \lambda_\alpha\ket{\phi_\alpha}\bra{\chi_\alpha}$, we have
\aln{
C(r)\ra\lrv{ \lrs{\frac{\lambda_{1}}{\lambda_0}}^r\braket{\chi_0|\sigma^z_{\tau=1}|\phi_1}\braket{\chi_1|\sigma^z_{\tau=1}|\phi_0}}
}
for $b\lesssim b_c$ and large $L$. Here, $0$ and $1$ respectively indicate the labels of eigenvalues with the largest and the second-largest modulus.}
From this the generalized correlation length is obtained as $\xi_\mr{cor}=-(\ln \lambda_1/\lambda_0)^{-1}\simeq \lambda_0/(\lambda_0-\lambda_1)$ and behaves as $\sim (b_c-b)^{-1/2}$ near the exceptional {DQPT}.

For $b>b_c$, $C(r)$ contains a term $e^{-ir\Delta}$ even in the thermodynamic limit, where $\Delta \:(<\pi)$ is the difference between angles of two complex-conjugate eigenvalues.
Thus the oscillation length becomes $\xi_\mr{osc}=\frac{2\pi}{\Delta}$ and behaves as $\sim (b-b_c)^{-1/2}$ near the exceptional {DQPT}.
\newline

\textbf{Partition-function zeros.}
Phase transitions occur when the zeros of the partition function  $e^{-LF_{L,T}}$, whose parameter (especially $b$ in our context) regime is extended to a complex one, accumulate at real values in the thermodynamic limit~\cite{Fisher78,Heyl18}.
Accumulation points of the partition-function zeros are thus read out from the points where maximum eigenvalues switch when we add proper perturbation $\delta b\: (\in\mbb{C})$ whose magnitude is infinitesimal~\cite{Andraschko14}.
Notably, the partition-function zeros accumulate along the real axis when the complex-conjugate pair contributes to maximum eigenvalues with $n_\mr{deg}= 2$  owing to AUS of $\tilde{U}$.
This is because one of the eigenvalues that form the complex conjugate at $b\in\mbb{R}$ becomes larger and smaller than the other for $b+\delta b$ and $b-\delta b\:(\delta b\in i\mbb{R})$, respectively.
Moreover, we find that these zeros on the real axis (say $b\geq b_c$) terminate at the exceptional {DQPT} ($b= b_c$).
This means that the exceptional {DQPT} corresponds to the realization of the edge singularity of the partition-function zeros at physical parameters on the real axis.
\newline

{\textbf{\large{Acknowledgments}}}
{
We are grateful to Kohei Kawabata, Nobuyuki Yoshioka, Keiji Saito, and Mamiko Tatsuta for fruitful comments.
We thank Jad C. Halimeh, Amit Dutta, and Subinay Dasgupta for notifying us of related papers.
}
The numerical calculations were carried out with the help of QUSPIN~\cite{Weinberg19}.
\newline

%\paragraph{Implication to classical dynamical phase transitions}
%\bibliographystyle{apsrev4-1}

%\bibliographystyle{naturemag}
%\bibliographystyle{nazo.bst}
%\bibliography{../refer_them}

%\end{comment}

\ 
\newline

\end{document}

\title{
 Supplementary Information for ``Exceptional Dynamical Quantum Phase Transitions in Periodically Driven Systems"
}
\author[1]{Ryusuke Hamazaki}
\affil[1]{Nonequilibrium Quantum Statistical Mechanics RIKEN Hakubi Research Team, RIKEN Cluster for Pioneering Research (CPR), RIKEN iTHEMS, Wako, Saitama 351-0198, Japan}

%\makeatletter
%\def\tagform@#1{\maketag@@@{(S-#1)}}
%\makeatother

%\renewcommand{\thefigure}{Supplementary Figure \arabic{figure}}

\renewcommand{\thetable}{Supplementary Equation \arabic{table}}

\renewcommand{\refname}{Supplementary References}

\maketitle

\section*{Supplementary Note 1: Dynamical phase transitions for other parameters and  initial/final states}

\begin{figure}
\begin{center}
\includegraphics[width=\linewidth]{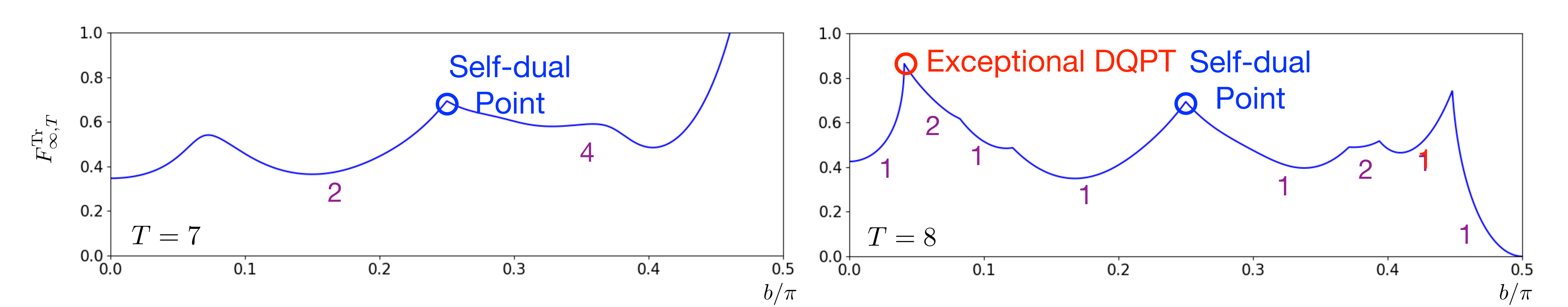}
\end{center}
\caption{Supplementary Figure 1. 
\textbf{(Real part of) dynamical free energy $F^\mr{Tr}_{\infty,T}$ as a function of $b$ for different transient times $T$.}
As varying $b$, the exceptional {dynamical quantum phase transition (DQPT)} occurs for $T=8$, which is prohibited for $T=7$.
Purple numbers denote $n_\mr{deg}$ for each phase.
We use $h=0.5$ and $J=0.25\pi$.
}
\label{Tchange}
\end{figure}

Here, we describe in detail dynamical phases and their transitions of the stroboscopic Ising model for situations different  from that presented in the main text.
Supplementary Figure~\ref{Tchange} shows examples of dynamical quantum phase transitions (DQPTs) of $F_\mr{\infty,T}^\mr{Tr}$ for $T=7$ and $T=8$ with $J=0.25\pi$ with varying $b$.
As discussed in the main text and  Supplementary Note 6,  $n_\mr{eff}$ [the number of degenerate eigenvalues with the largest modulus of $\tilde{U}$] is always equal to or larger than 2 for $T=7$ because the system belongs to Class AII ($n_\mr{eff}\geq 4$ comes from additional symmetry $\tilde{U}$, such as the translation invariance).
Therefore no exceptional DQPT occurs in this case.
On the other hand, we find the exceptional DQPT for $T=8$, indicating that the exceptional DQPT is the general mechanism that can occur in systems with Class AI-type antiunitary symmetry.

\begin{figure}
\begin{center}
\includegraphics[width=\linewidth]{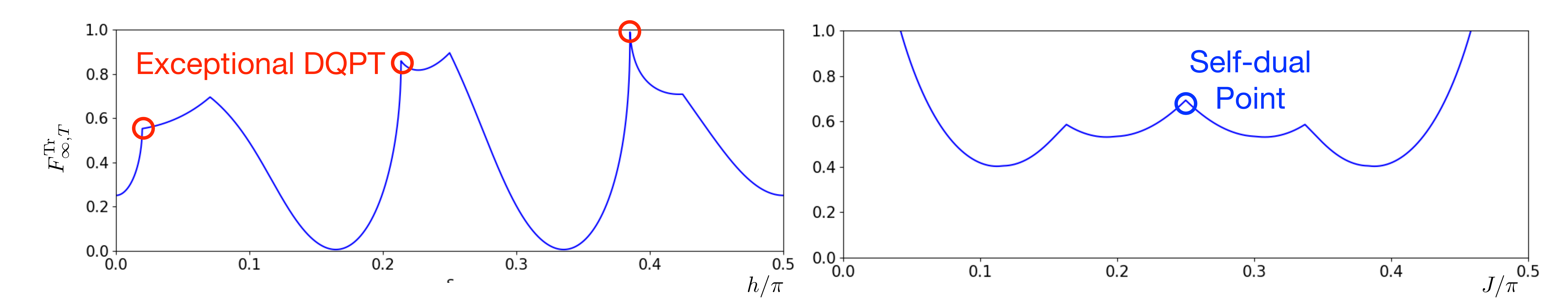}
\end{center}
\caption{Supplementary Figure 2.
\textbf{(Real part of) dynamical free energy $F^\mr{Tr}_{\infty,T}$ as a function of $h$ and $J$.}
As a function of $h$, we have several exceptional {dynamical quantum phase transitions (DQPTs)}, where $J=0.25\pi$, $b=0.05\pi$ and $T=6$ are used.
On the other hand, no exceptional DQPT exists as a function of $J$ because of the absence of the hidden antiunitary symmetry of $\tilde{U}$ ($h=3$, $b=-0.25\pi$ and $T=6$ are used).
}
\label{hJchange}
\end{figure}

Next, Supplementary Figure~\ref{hJchange} shows the case where we vary parameters other than $b$. We find that the exceptional DQPT can occur for $J=0.25\pi$ even when $h$ is varied for fixed appropriate $b$, which indicates the divergence of the generalized observable $\braket{\sigma_1^z}_\mr{gexp}$.
On the other hand, the exceptional DQPT does not occur if $J$ is varied since antiunitary symmetry no longer exists in $\tilde{U}$.

\begin{figure}
\begin{center}
\includegraphics[width=\linewidth]{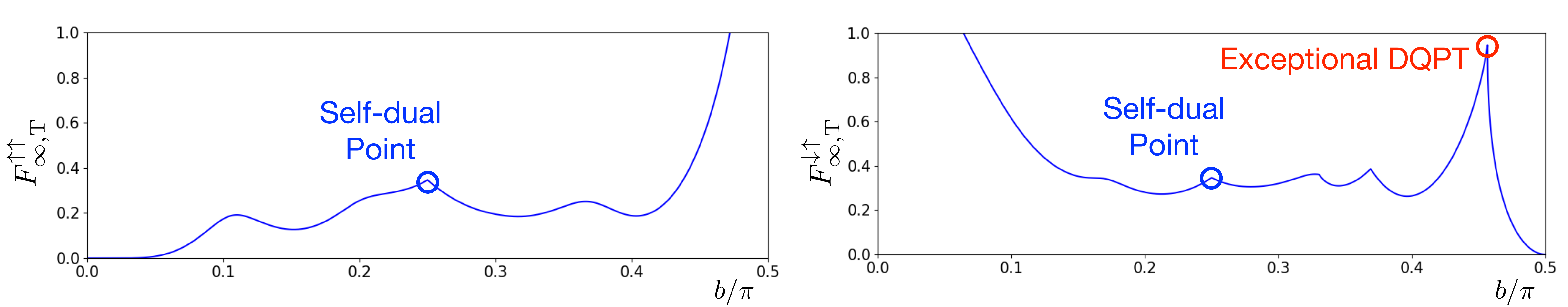}
\end{center}
\caption{Supplementary Figure 3.
\textbf{(Real part of) dynamical free energy $F^{\uparrow\uparrow}_{\infty,T}$ and  $F^{\downarrow\uparrow}_{\infty,T}$ as a function of $b$.}
The exceptional {dynamical quantum phase transition (DQPT)} occurs only for $F^{\downarrow\uparrow}_{\infty,T}$. Note that $n_\mr{deg}$ changes from 2 to 1 at the exceptional DQPT as increasing $b$.
We use $h=2, J=-0.25\pi$ and $T=7$ for both of the data.
}
\label{inichange}
\end{figure}

Finally, Supplementary Figure~\ref{inichange} shows the $b$-dependence of $F_{\infty,T}^{\uparrow\uparrow}$
and $F_{\infty,T}^{\downarrow\uparrow}$ for $J=0.25\pi$ and $T=7$.
As detailed in  Supplementary Note 3, $\tilde{U}_\mr{\downarrow\uparrow}$ has Class AI antiunitary symmetry for odd $T$, but  $\tilde{U}_\mr{\uparrow\uparrow}$ does not.
Thus, while we can find the exceptional DQPT for $F_{\infty,T}^{\downarrow\uparrow}$ but not for $F_{\infty,T}^{\uparrow\uparrow}$.

\section*{Supplementary Note 2: Derivation of spacetime-dual operators}
Here, we describe in detail the derivation of the spacetime-dual operators to calculate the dynamical free energies, following Refs.~\cite{Akila16,Bertini18,Bertini19}.
We first seek for the representation for $\tilde{U}_\mr{Tr}$, which satisfies
\aln{\label{jouken}
\frac{\Tr[U^T]}{2^L}=\Tr[\tilde{U}_\mr{Tr}^L]
}
or equivalently
\aln{
F^\mr{Tr}_{L,T}=-\frac{\log|\Tr[\tilde{U}_\mr{Tr}^L]|}{L}.
}
For this purpose, we notice
\aln{
\Tr[U^T]&=\sum_\lrm{\mbf{s}_\tau}\prod_{\tau=1}^T\braket{\mbf{s}_{\tau+1}|e^{-i\sum_{j=1}^L b\sigma_j^x}e^{-i\sum_{j=1}^LJ\sigma_j^z\sigma_{j+1}^z-i\sum_{j=1}^Lh\sigma_j^z}|\mbf{s}_\tau}\nonumber\\
&=\lrs{\frac{\sin 2b}{2i}}^{LT/2}\sum_\lrm{s_{\tau,j}}
e^{-i\sum_{\tau=1}^T\sum_{j=1}^L(J s_{\tau, j} s_{\tau, j+1}+J^{\prime} s_{\tau, j} s_{\tau+1, j}+h s_{\tau, j})},
}
where $\ket{\mbf{s}_\tau}$ are the computational basis, $s_{\tau,j}$ are classical spin variables taking $\pm 1$, and $J'=-\frac{\pi}{4}-\frac{i}{2}\log\tan b$.
On the other hand, we can consider 
\aln{
\tilde{U}'_\mr{Tr}=e^{-i\sum_{\tau=1}^T \tilde{b}\sigma^x_\tau}e^{-i\sum_{\tau=1}^T (\tilde{J}\sigma^z_\tau\sigma^z_{\tau+1}+{h}\sigma_\tau^z)},
}
which satisfies
\aln{
\Tr[\tilde{U}'^L_\mr{Tr}]
=\lrs{\frac{\sin 2\tilde{b}}{2i}}^{LT/2}\sum_\lrm{s_{\tau,j}}
e^{-i\sum_{\tau=1}^T\sum_{j=1}^L(\tilde{J}' s_{\tau, j} s_{\tau, j+1}+\tilde{J} s_{\tau, j} s_{\tau+1, j}+h s_{\tau, j})}
}
with $\tilde{J}'=-\frac{\pi}{4}-\frac{i}{2}\log\tan \tilde{b}$.
Then, introducing a normalization constant 
\aln{
C=\frac{1}{2}\lrs{\frac{\sin 2b}{\sin 2\tilde{b}}}^{\frac{LT}{2}}
}
and setting $\tilde{b}=\mr{arctan}[e^{2i(J+\pi/4)}]=\frac{i}{2}\log\lrs{\frac{1+e^{2iJ}}{1-e^{2iJ}}}=-\frac{\pi}{4}-\frac{i}{2}\log\tan J$ (to satisfy $\tilde{J}'=J$) with $\tilde{J}=J'$,
 we have 
 \aln{
 \tilde{U}_\mr{Tr}=C\tilde{U}'_\mr{Tr}=Ce^{-i\sum_{\tau=1}^T \tilde{b}\sigma^x_\tau}e^{-i\sum_{\tau=1}^T (\tilde{J}\sigma^z_\tau\sigma^z_{\tau+1}+{h}\sigma_\tau^z)},
 }
which satisfies Supplementary Equation~\eqref{jouken}
as desired.

Next, $\tilde{U}_{\uparrow\uparrow/\downarrow\uparrow}$ can be calculated similarly.
We show that they are represented as 
\aln{\label{open}
\tilde{U}_{\uparrow\uparrow/\downarrow\uparrow}=C'e^{-i\sum_{\tau=1}^{T-1} \tilde{b}\sigma_\tau^x}e^{-i\sum_{\tau=1}^{T-2}\tilde{J}\sigma_\tau^z\sigma_{\tau+1}^z-i\sum_{\tau=1}^{T-1}h\sigma_\tau^z-i\tilde{J}(\sigma_1^z+I\sigma_{T-1}^z)}
}
with the open boundary condition, where 
$C'=(\sin 2b/2i)^{1/2}(\sin 2b/\sin 2\tilde{b})^{(T-1)/2}e^{-i(h+J)}$
, and $I=1\:(-1)$ for $\tilde{U}_{\uparrow\uparrow}\:(\tilde{U}_{\downarrow\uparrow})$.
To see this, we notice (for $\ket{\psi_f}=\ket{\uparrow\cdots\uparrow}/\ket{\downarrow\cdots\downarrow}$)
\aln{
\braket{\psi_f|U^T|\uparrow\cdots\uparrow}=&
\sum_{\lrm{\mbf{s}_\tau}}\braket{\psi_f|U|\mbf{s}_{T-1}}\cdots\braket{\mbf{s}_1|U|\uparrow\cdots\uparrow}\nonumber\\
=\lrs{\frac{\sin 2b}{2i}}^{LT/2}\sum_{\lrm{s_{\tau,j}}}&e^{-i\sum_{\tau=1}^{T-1}\sum_{j=1}^LJs_{\tau,j}s_{\tau,j+1}+hs_{\tau,j}-i\sum_{\tau=1}^{T-2}\sum_{j=1}^LJ's_{\tau,j}s_{\tau+1,j}}\nonumber\\
&\times e^{-i\sum_{j=1}^L\lrm{(J+h)+J's_{1,j}+J'Is_{T-1,j} }},
}
where $I=1$ for $\ket{\psi_f}=\ket{\uparrow\cdots\uparrow}$ and $I=-1$ for $\ket{\psi_f}=\ket{\downarrow\cdots\downarrow}$.
To  construct dual operators, we consider $(T-1)$-spins along time with the open boundary condition.
In fact, if we assume Supplementary Equation~\eqref{open}, we find
\aln{
F^{\uparrow\uparrow/\downarrow\uparrow}_{L,T}=-\frac{\log|\Tr[\tilde{U}_{\uparrow\uparrow/\downarrow\uparrow}^L]|}{L}.
}

\section*{Supplementary Note 3: Existence of antiunitary symmetry}
Here, we describe in detail the antiunitary symmetry (AUS) of the spacetime-dual operator.
First, we show that the spacetime-dual operator $\tilde{U}_\mr{Tr}$ with $J=\frac{\pi}{4}+\frac{n\pi}{2}\:(n\in\mbb{Z})$ satisfies 
\aln{\label{symtr}
V\tilde{U}_\mr{Tr}^* V^\dag=e^{i\phi}\tilde{U}_\mr{Tr},
}
where
\aln{
V=\prod_{\tau=1}^Te^{i\frac{\pi}{2}\sigma_\tau^y}
}
and $\phi\in\mbb{R}$.
%Note that  $V$ becomes TRS for $\phi=0$ and   the conjugate of particle-hole symmetry (PHS$^\dag$) for $\phi=\pi$~\cite{Kawabata19S}.
%On the other hand, symmetries parametrized by $\phi$ (including TRS and PHS$^\dag$) are equivalent in a sense that the phase can be eliminated by the transformation $\tilde{U}\ra e^{-i\phi/2}\tilde{U}$~\cite{Kawabata18T}.
%We thus call symmetries satisfying Supplementary Equation~\eqref{symtr} simply TRS.

%We show Supplementary Equation~\eqref{symtr} for $J=-\pi/4$ for simplicity (the derivation is essentially the same for other $J$).
For $J=\frac{\pi}{4}+\frac{n\pi}{2}\:(n\in\mbb{Z})$, $\tilde{b}=\pm\pi/4$ becomes real.
Noticing $\tilde{J}=-\frac{\pi}{4}-\frac{i}{2}\log\tan b=-\tilde{J}^*-\frac{\pi}{2}$, the left-hand side of Supplementary Equation~\eqref{symtr}  becomes
\aln{
V\tilde{U}_\mr{Tr}^* V^\dag&=CVe^{i\sum_{\tau=1}^T \tilde{b}\sigma^x_\tau}V^\dag Ve^{i\sum_{\tau=1}^T (\tilde{J}^*\sigma^z_\tau\sigma^z_{\tau+1}+{h}\sigma_\tau^z)}V^\dag\nonumber\\
&=e^{-i\sum_{\tau=1}^T \tilde{b}\sigma^x_\tau}e^{i\sum_{\tau=1}^T (\tilde{J}^*\sigma^z_\tau\sigma^z_{\tau+1}-{h}\sigma_\tau^z)}\nonumber\\
&=e^{-i\sum_{\tau=1}^T \tilde{b}\sigma^x_\tau}e^{-i\sum_{\tau=1}^T (\tilde{J}\sigma^z_\tau\sigma^z_{\tau+1}+{h}\sigma_\tau^z)}e^{-i\frac{\pi}{2}\sum_{\tau=1}^T\sigma^z_\tau\sigma^z_{\tau+1}}\nonumber\\
&=\tilde{U}_\mr{Tr}\prod_{\tau=1}^T\lrs{-i\sigma_\tau^z\sigma_{\tau+1}^z}=(-i)^T\tilde{U}_\mr{Tr},
}
which is the right-hand side.
Since $VV^*=\mbb{I}$ for even $T$ and $-\mbb{I}$ for odd $T$, $\tilde{U}_\mr{Tr}$ belongs to Class AI for even $T$ and Class AII for odd $T$.

Next, we show that $\tilde{U}_{\downarrow\uparrow}$ with $J=\frac{\pi}{4}+\frac{n\pi}{2}\:(n\in\mbb{Z})$ satisfies 
\aln{\label{symtr2}
V\tilde{U}_{\downarrow\uparrow}^* V^\dag=e^{i\phi}\tilde{U}_{\downarrow\uparrow},
}
where
\aln{
V=\mc{P}\prod_{\tau=1}^{T-1}e^{i\frac{\pi}{2}\sigma_\tau^y}
}
and $\phi\in\mbb{R}$. Here, $\mc{P}$ is the parity operator exchanging the site $\tau$ and $T-\tau$.
In fact, the left-hand side of Supplementary Equation~\eqref{symtr2} becomes
\aln{
V\tilde{U}_{\downarrow\uparrow}^* V^\dag&=
C'^*Ve^{i\sum_{\tau=1}^{T-1} \tilde{b}\sigma_\tau^x}V^\dag Ve^{i\sum_{\tau=1}^{T-2}\tilde{J}^*\sigma_\tau^z\sigma_{\tau+1}^z+i\sum_{\tau=1}^{T-1}h\sigma_\tau^z+i\tilde{J}^*(\sigma_1^z-\sigma_{T-1}^z)}V^\dag\nonumber\\
&=e^{i\zeta}C'e^{-i\sum_{\tau=1}^{T-1} \tilde{b}\sigma_\tau^x}
e^{i\sum_{\tau=1}^{T-2}\tilde{J}^*\sigma_\tau^z\sigma_{\tau+1}^z-i\sum_{\tau=1}^{T-1}h\sigma_\tau^z-i\tilde{J}^*(-\sigma_1^z+\sigma_{T-1}^z)}\nonumber\\
&=e^{i\zeta}\tilde{U}_{\downarrow\uparrow}e^{-i\frac{\pi}{2}\sum_{\tau=1}^{T-1}\sigma_\tau^z\sigma_{\tau+1}^z+i\frac{\pi}{2}(-\sigma_1^z+\sigma_{T-1}^z)}\nonumber\\
&=e^{i\zeta}\tilde{U}_{\downarrow\uparrow}\sigma_1^z\sigma_{T-1}^z\prod_{\tau=1}^{T-1}\lrs{-i\sigma_\tau^z\sigma_{\tau+1}^z}\nonumber\\
&=e^{i\zeta}(-i)^{T-1}\tilde{U}_{\downarrow\uparrow},
}
which is the right-hand side.
Here, we have used $\tilde{b}\in\mbb{R}$ for $J=\frac{\pi}{4}+\frac{n\pi}{2}\:(n\in\mbb{Z})$, $\tilde{J}^*=-(\tilde{J}+\pi/2)$, 
$C'^*/C'=e^{i\zeta}\:(\zeta\in\mbb{R})$, and $\mc{P}\sigma_1\mc{P}=\sigma_{T-1}$.
Since $VV^*=\mbb{I}$ for odd $T$ and $-\mbb{I}$ for even $T$, $\tilde{U}_{\downarrow\uparrow}$ belongs to Class AI for odd $T$ and Class AII for even $T$.

We note that the minus sign associated with the exchange between $\sigma_1^z-\sigma_{T-1}^z$ under parity operation is essential for this antiunitary symmetry.
This is not possible for $\tilde{U}_{\uparrow\uparrow}$, where 
$\sigma_1^z+\sigma_{T-1}^z$ is invariant under the parity operation.

 {
\section*{Supplementary Note 4: Thermalization of the expectation values of local observables averaged over a long time}
As mentioned in the main text,
 DQPTs in our model do not appear as an infinite-time average of expectation values of local observables because of the Floquet eigenstate thermalization hypothesis~\cite{Kim14}.
To demonstrate this, here we numerically show that  the time-averaged expectation values of local observables become the thermalized values.
}
 
  {
We particularly consider a time-averaged expectation value of local magnetization
\aln{\label{tavelm}
\av{m^z}=\frac{1}{T}\sum_{t=1}^T\braket{\psi_i|U^{-t}\sigma_{i=1}^zU^t|\psi_i},
}
and that of local correlation
\aln{\label{tavelc}
\av{C^{zz}}=\frac{1}{T}\sum_{t=1}^T\braket{\psi_i|U^{-t}\sigma_{i=1}^z\sigma_{i=2}^zU^t|\psi_i},
}
where $T$ is sufficiently large.
If the system thermalizes, $\av{m^z}$ and $\av{C^{zz}}$ will be equal to the expectation value at the infinite temperature, i.e., $\Tr[\sigma_{i=1}^z]/2^L=\Tr[\sigma_{i=1}^z\sigma_{i=2}^z]/2^L=0$.
In the following, we take an initial state as $\ket{\psi_i}=\ket{\uparrow\uparrow\cdots \uparrow}$.
We remind that the DQPTs (including the exceptional DQPT) occur for several $b$ for $F_{\infty,T}^{\lrm{\uparrow\uparrow/\downarrow\uparrow}}$ with finite $T$ starting from this initial state (see  Supplementary Note 1).
}

 {
Supplementary Figure \ref{longexp} shows the values of $\av{m^z}$ and $\av{C^{zz}}$ as a function of $b$ for different system sizes $L$.
We find that $\av{m^z}$ and $\av{C^{zz}}$ approach zero as increasing the system size especially for $b$ far from the integrable point ($b/\pi=0, 0.5, 1$), which indicates that they become zero in the thermodynamic limit.
This implies that our DQPTs are unique to finite-time regimes, in which time serves as an important parameter in stark contrast with conventional phase transitions.
}

 {
\begin{figure}
\begin{center}
\includegraphics[width=\linewidth]{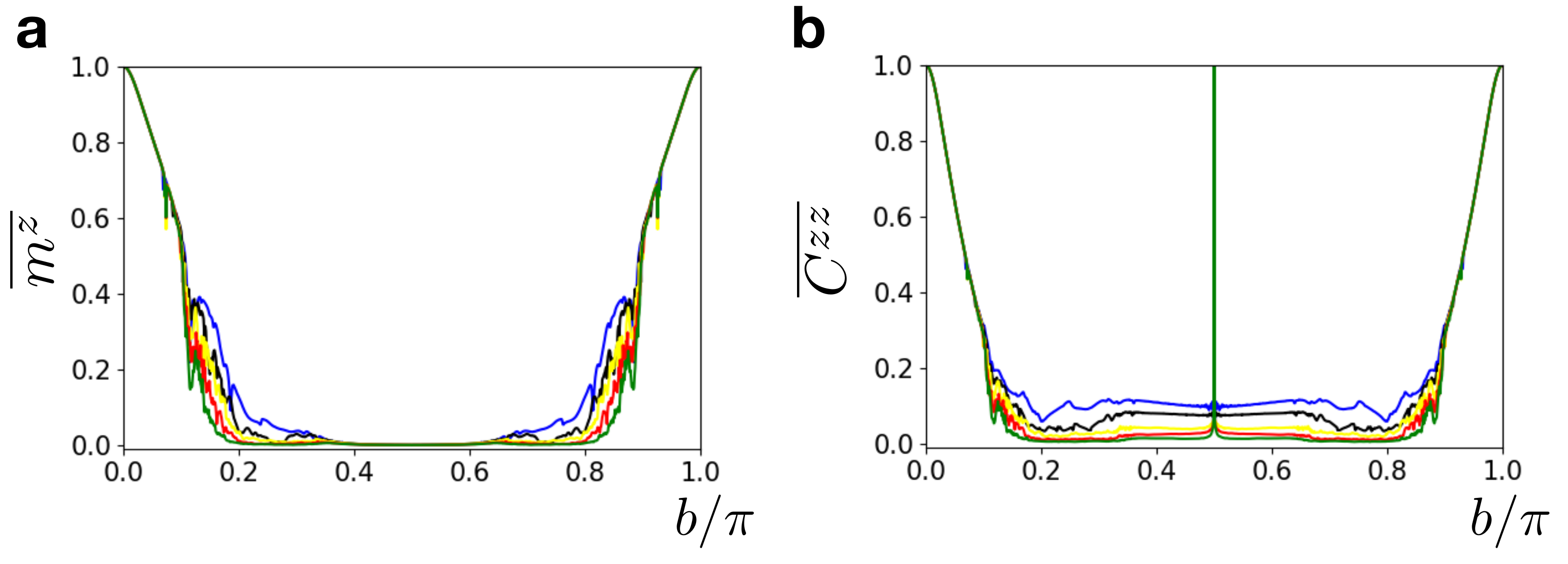}
\end{center}
\caption{Supplementary Figure 4.
\textbf{Long time average of the expectation values of local observables in \textbf{a.} Supplementary Equation~\eqref{tavelm}  and \textbf{b.} Supplementary Equation~\eqref{tavelc} as a function of $b$.}
We show results for different system sizes, $L=8$ (blue), 10 (black) 12 (yellow), 14 (red), and 16 (green).
We find that  $\av{m^z}$ and $\av{C^{zz}}$ approach zero as increasing the system size especially for $b$ far from the integrable point ($b/\pi=0, 0.5, 1$), which indicates that they become zero in the thermodynamic limit.
This means that, while we have several {dynamical quantum phase transitions} of $F_{\infty,T}^{\lrm{\uparrow\uparrow/\downarrow\uparrow}}$ with finite $T$ for the displayed range of $b$, long-time-averages of the expectation values thermalize to the values described by the infinite-temperature state.
We use $T=1000$, $J=-0,25\pi$ and $h=1.3\pi$.
}
\label{longexp}
\end{figure}
}

 {
\section*{Supplementary Note 5: Dynamical phase transition at the self-dual point}
In this section, we discuss the details for the DQPT occurring at the self-dual points, i.e., $J=\frac{\pi}{4}+\frac{n\pi}{2}$ and $b=\frac{\pi}{4}+\frac{m\pi}{2}\:(n,m\in\mbb{Z})$~\cite{Bertini18,Bertini19}.
We especially show that crossing self-dual points entails DQPTs universally for $F^{\mr{Tr}/\uparrow\uparrow/\downarrow\uparrow}_{\infty,T}$ with any $T$ and $h$, whose singularity is analogous to that for the conventional DQPT.
Moreover, the dynamical free energy density takes a universal value as $F^\mr{Tr}_{\infty,T}=\log 2$ or $F^{\uparrow\uparrow/\downarrow\uparrow}_{\infty,T}=\log 2/2$ there.
}

 {
We especially focus on the case  for $F_\mr{Tr}$ (other dynamical free energies are discussed in a similar manner).
As discussed in the main text,
 we can exactly write 
\aln{
\tilde{U}_\mr{Tr}=Ce^{-i\sum_{\tau=1}^T \tilde{b}\sigma_\tau^x}e^{-i\sum_{\tau=1}^T\tilde{J}\sigma_\tau^z\sigma_{\tau+1}^z-i\sum_{\tau=1}^Th\sigma_\tau^z}
}
using the spacetime duality~\cite{Akila16,Bertini18,Bertini19}.
Here, $\tilde{b}=-\pi/4-i\log(\tan J)/2$, $\tilde{J}=-\pi/4-i\log(\tan b)/2$ and $C=(\sin 2b/\sin 2\tilde{b})^{T/2}/2$.
Importantly, $\tilde{U}$ is unitary (up to a constant) only at the self-dual points.
}

 {
As noted in the main text, for typical cases, DQPTs occur when maximum of two eigenvalues with different $\theta_\alpha$ switches accidentally, where $n_\mr{deg}=1$ for each phase and $n_\mr{deg}=2$  at transition  (Supplementary Figure~\ref{eigenvalue}\textbf{a}).
On the other hand, DQPT occurs more universally if the self-dual point is crossed:
all of the modulus of the eigenvalues of $\tilde{U}$ are equal at this point ($n_\mr{deg}=2^T$),
and crossing this point typically switches the largest eigenvalue (Supplementary Figure~\ref{eigenvalue}\textbf{b}).
At transition, dynamical free energies are determined only by the modulus of the eigenvalues, which leads to universal values $F^\mr{Tr}_{\infty,T}=\log 2$ (similarly, we have $F^{\uparrow\uparrow/\downarrow\uparrow}_{\infty,T}=\log 2/2$).
To our best knowledge, this is the first evidence that the self-dual point is a critical point of different dynamical phases.
}

 {
\begin{figure}
\begin{center}
\includegraphics[width=\linewidth]{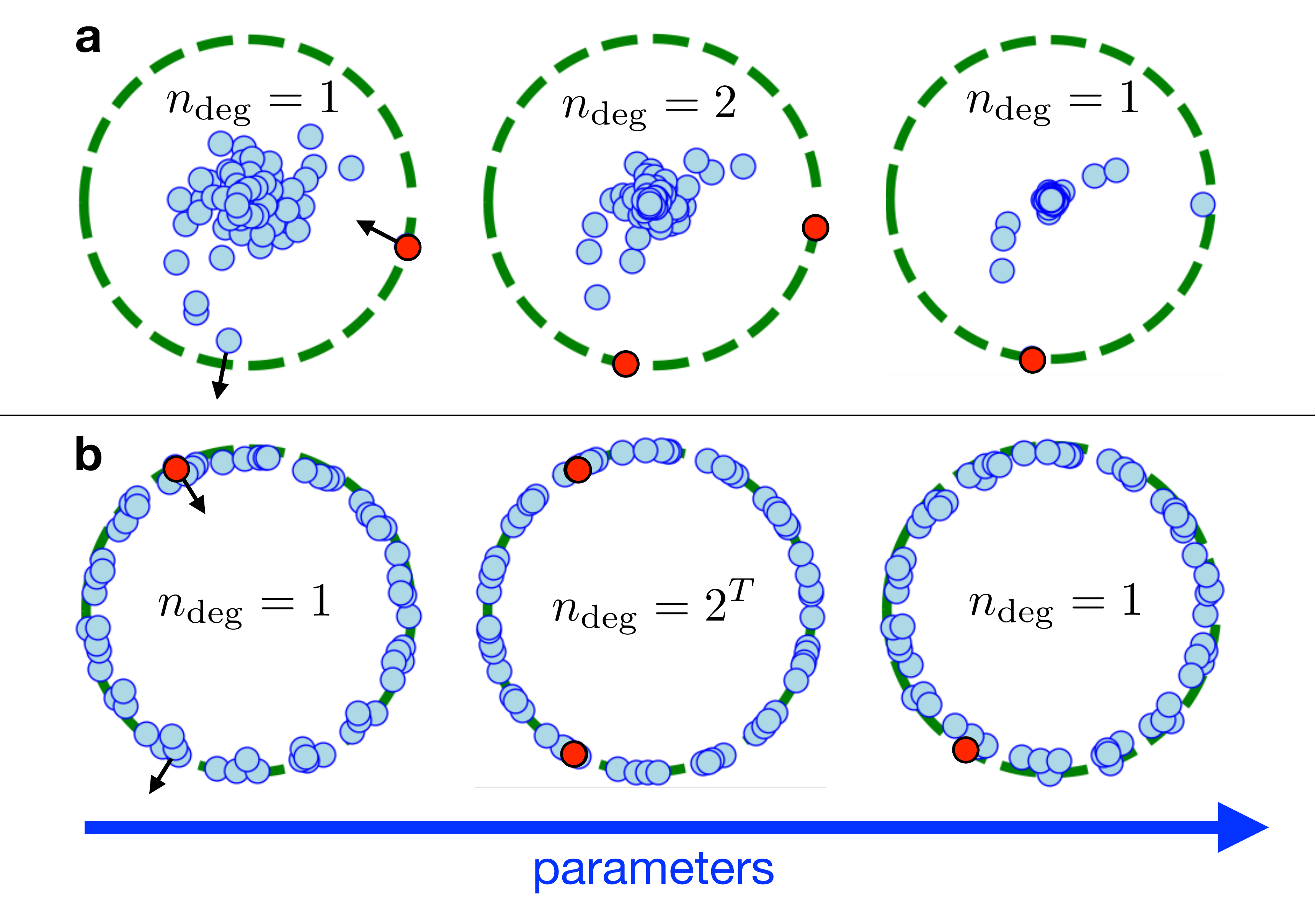}
\end{center}
\caption{Supplementary Figure 5.
\textbf{Schematic of eigenvalue dynamics of the spacetime-dual operator $\tilde{U}$ and its relation to the dynamical quantum phase transition (DQPT).}
\textbf{a.} Typical eigenvalue dynamics  (small circles) near DQPT.
Green dashed circles have the radius that corresponds to the eigenvalue(s) with the largest modulus.
The eigenvalue with the largest modulus (red circles) switches at the critical point, at which two eigenvalues have the same modulus $n_\mr{deg}=2$.
\textbf{b.} Eigenvalue dynamics through the DQPT induced by self-dual points.
At the self-dual point, all of $2^T$ numbers of eigenvalues have the same modulus owing to the unitarity of $\tilde{U}$.
}
\label{eigenvalue}
\end{figure}
}

 {
\section*{Supplementary Note 6: Dynamical phase transitions in finite systems}
Here, we show several results concerning the DQPTs for finite system size $L$.
}

 {
\subsection*{Degeneracy and finite-size effect}
While the number of eigenvalues with maximum modulus $n_\mr{deg}$ does not contribute to the free energy density for $L\ra\infty$, $n_\mr{deg}$ can characterize each phase via a finite-size correction $\Delta F_{L,T}=F_{L,T}-F_{\infty,T}$.
Indeed, the second term in 
\aln{\label{eq:fltdual}
F_{L,T}\simeq -\log |\lambda_\mr{M}|-\frac{1}{L}\log\lrv{\sum_\alpha e^{i\theta_\alpha L}}.
}
shows that
$\Delta F_{L,T}$ is upper bounded by $\log n_\mr{deg}/L$, where the bound is achieved when $e^{i\theta_\alpha L}=e^{i\theta_\beta L}$ for every $\alpha\neq\beta$.
}

 {
Supplementary Figure~\ref{scaling}\textbf{a} shows the finite-size scaling of $\Delta F_{L,T}^\mr{Tr}$ near the exceptional DQPT. 
While $\Delta F_{L,T}^\mr{Tr}$ exponentially decays with $L$ for $b<b_c$ since $n_\mr{deg}=1$, it exhibits polynomially decaying oscillations for $b>b_c$, where the decay is upper bounded by $\log n_\mr{deg}/L=\log 2/L$ for sufficiently large $L$.
At the transition point, oscillation-free polynomial decay is observed.
Another example is the behavior at the self-dual point, where the decay is upper bounded by $\log 2^T/L$ (Supplementary Figure~\ref{scaling}\textbf{b}).
}

 {
\begin{figure}
\begin{center}
\includegraphics[width=\linewidth]{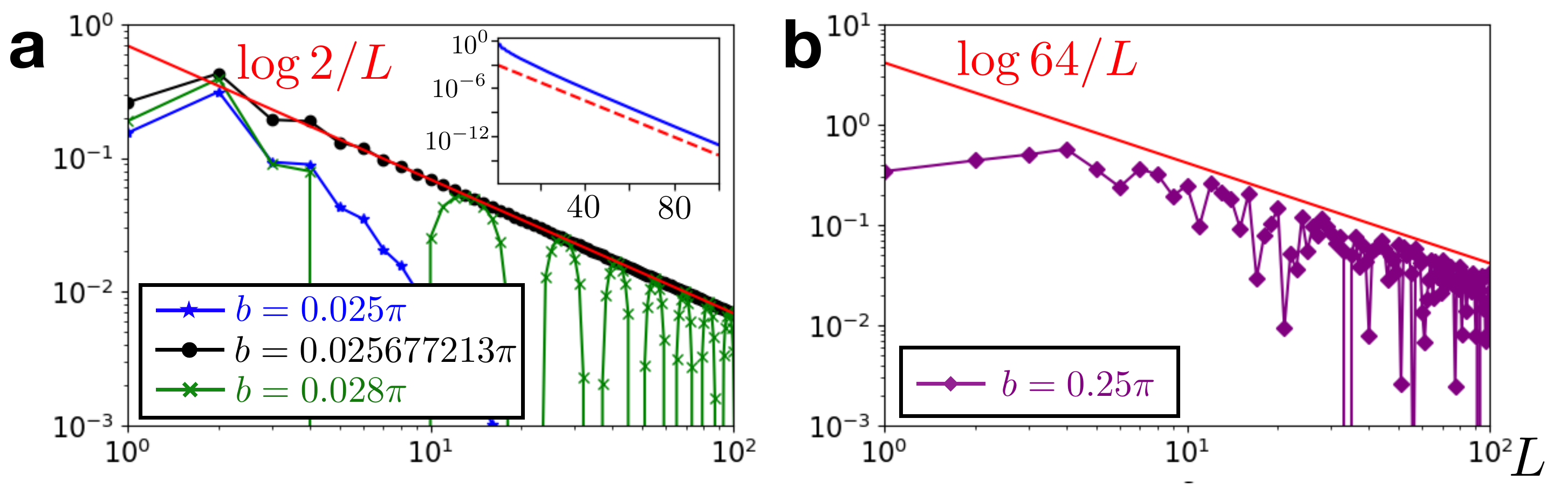}
\end{center}
\caption{Supplementary Figure 6.
\textbf{Log-log plot for finite-size correction of the dynamical free energy density $\Delta F_{L,T}^\mr{Tr}$.}
\textbf{a.} Behavior near the exceptional dynamical quantum phase transition (DQPT). Below the DQPT with $n_\mr{deg}=1$ (blue), $\Delta F_{L,T}^\mr{Tr}$ decays rapidly. (inset) Semi-log plot of the same data shows that it indeed decays exponentially (dotted red line is an eyeguide for exponential decay).
Above the DQPT with $n_\mr{deg}=2$ (green), $\Delta F_{L,T}^\mr{Tr}$ decays with oscillations.
The decay is bounded by $\log 2/L$ (red solid line) for sufficiently large $L$.
Approaching the critical point (black), the oscillation vanishes and $\Delta F_{L,T}^\mr{Tr}\sim \log2/L$.
\textbf{b.} Behavior at the self-dual point (purple), where $n_\mr{deg}=2^T=64$.
The correction is bounded by $\log 64/L$.
We use $J=-\pi/4$ and $h=3.0$, and $T=6$.
}
\label{scaling}
\end{figure}
}

 {
The number $n_\mr{deg}$  thus provides universal information through $\Delta F_{L,T}$ on dynamical phases, which is deeply related to the symmetries hidden in the space-time dual operator $\tilde{U}$.
If $\tilde{U}$ is in Class AI ($F^\mr{Tr}_{L,T}$ with even $T$ and  $F^{\downarrow\uparrow}_{L,T}$ with odd $T$ under the condition $J=-\pi/4+n\pi/2\:(n\in\mbb{Z})$), phases with $n_\mr{deg}=2$ can naturally appear as well as phases with $n_\mr{deg}=1$.
If $\tilde{U}$ is in Class AII ($F^\mr{Tr}_{L,T}$ with odd $T$ and  $F^{\downarrow\uparrow}_{L,T}$ with even $T$ under the condition $J=-\pi/4+n\pi/2\:(n\in\mbb{Z})$), all phases satisfy $n_\mr{deg}\geq 2$. 
We note that other symmetries are found to exist that enhance the value of $n_\mr{deg}$, such as translation invariance for $\tilde{U}_\mr{Tr}$ or integrability at $h=0$, which may also be interesting to study systematically.
}

\subsection*{Generalized correlation function in small systems}
While the true DQPT occurs in the thermodynamic limit, qualitative signature of AUS-unbroken and AUS-broken phases are already captured even for finite system sizes with the generalized correlation function.
In Supplementary Figure~\ref{smallcorrelation}, we show the (normalized) generalized correlation function for different system size $L$.
Even for $L=10$, which has been prepared in experiments of trapped ions~\cite{Zhang17}, we find clear difference between AUS-unbroken ($b<b_c\simeq 0.0257\pi$) and AUS-broken ($b>b_c$) regime.
Indeed, $C(r)$ decays fast for the former but does not decay for the latter.
If we consider $L=20$, we can also see the oscillatory behavior in the AUS-broken regime:
note that the oscillation length is evaluated as $\xi_\mr{osc}\simeq 10.4$ for $b=0.03\pi$ and 
$\xi_\mr{osc}\simeq 8.5$  for $b=0.0325\pi$.

\begin{figure}
\begin{center}
\includegraphics[width=\linewidth]{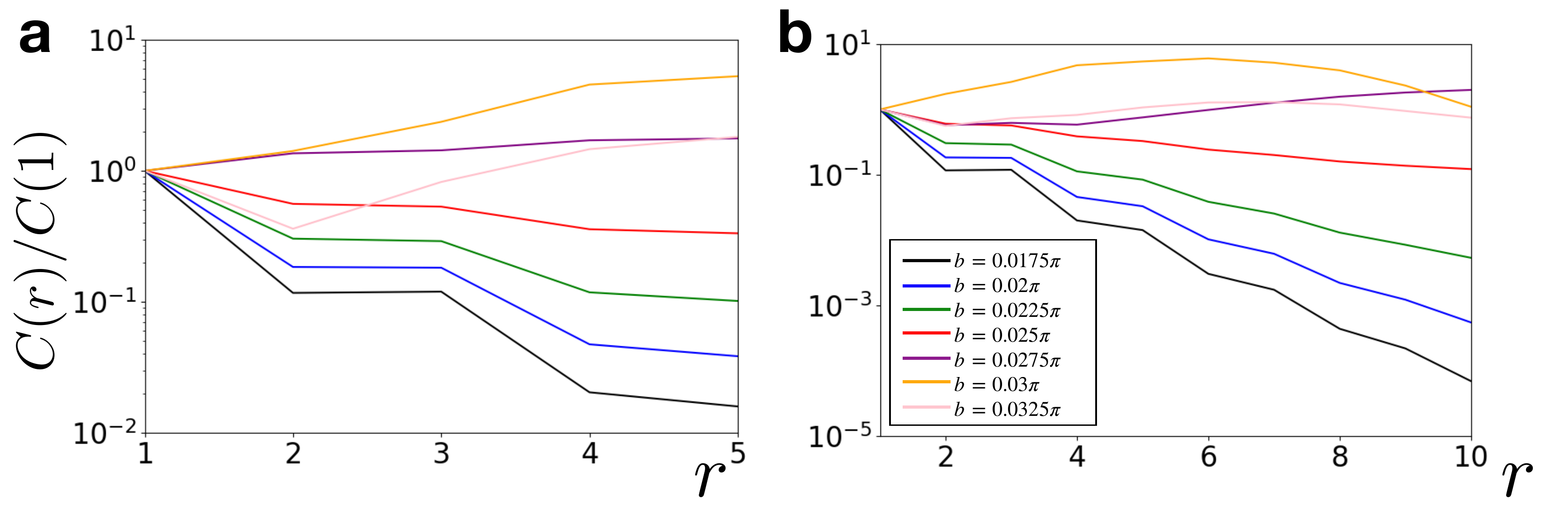}
\end{center}
\caption{Supplementary Figure 7.
\textbf{Normalized generalized correlation function $C(r)/C(0)$ for different values of $b$.}
\textbf{a.} Results for $L=10$.
Clear difference between antiunitary-symmetry (AUS) unbroken ($b<b_c\simeq 0.0257\pi$) and AUS broken ($b>b_c$) regimes already appears.
Indeed, $C(r)$ decays fast for the former but does not decay for the latter.
\textbf{b.} Results for $L=20$.
The oscillatory behavior in the AUS-broken regime arises,
where the oscillation length is evaluated as $\xi_\mr{osc}\simeq 10.4$ for $b=0.03\pi$ and 
$\xi_\mr{osc}\simeq 8.5$  for $b=0.0325\pi$.
We use $J=-\pi/4$ and $h=3.0$, and $T=6$.
}
\label{smallcorrelation}
\end{figure}

 {
\section*{Supplementary Note 7: Explicit construction of an order parameter}
We here show that we can explicitly construct an order parameter using different-time generalized observables.
For this purpose, we especially focus on $F^{\downarrow\uparrow}_{L,T}=-\frac{1}{L}\log|\braket{\psi_f|U^T|\psi_i}|$ with $\ket{\psi_i}=	\ket{\uparrow\cdots\uparrow}$ and $\ket{\psi_f}=\ket{\downarrow\cdots\downarrow}$, instead of $F^{\Tr}_{L,T}$, since its operational meaning in  experimental situations is more direct.
We note that $F^{\downarrow\uparrow}_{\infty,T}$ shows the exceptional DQPT for $b=b_c\simeq 0.446\pi$ with $h=1.3$, $T=5$ and $J=-\pi/4$, where the AUS is broken for $b<b_c$ and unbroken for $b>b_c$ (note that this is opposite to the case for $F^\mr{Tr}_{\infty,T}$).
}

 {
For our discussion, we first note that, for a usual symmetry breaking, such as $\mbb{Z}_2$ symmetry breaking of an Ising model, magnetization $m^z$ becomes an order parameter.
In this case, $m^z$ is odd under symmetry operation ($m^z\ra -m^z$), so $\braket{m^z}=-\braket{m^z}=0$ when symmetry of the state is unbroken.
}

 {
Similarly, our antiunitary symmetry operation in the space-time dual space ($V=\mathcal{P} \prod_{\tau=1}^{T-1} e^{i \frac{\pi}{2} \sigma_{\tau}^{y}}$ for $\tilde{U}_{\downarrow\uparrow}$) 
is found to correspond to a combined symmetry of time exchange $\tau\leftrightarrow T-\tau$, complex conjugation, and certain spin reversal in  the original space. 
For example,  using the spacetime-dual transformation and the spectral decomposition $\tilde{U}_{\downarrow\uparrow}=\sum_\alpha \lambda_\alpha\ket{\phi_\alpha}\bra{\chi_\alpha}$, we have
\aln{
\braket{\psi_f|U^{T-\tau}\sigma_i^zU^{\tau}|\psi_i}&=\Tr[\tilde{U}_{\downarrow\uparrow}^L\sigma_\tau^z]\\
&\ra \lambda_0^L\braket{\chi_0|\sigma_\tau^z|\phi_0}
}
for the symmetry-unbroken phase, where 0 labels the eigenvalues with the largest modulus.
In this phase, we have
\aln{
V\ket{\phi_0}^*=\ket{\phi_0},\quad V\ket{\chi_0}^*=\ket{\chi_0},
}
and thus
\aln{
\braket{\chi_0|\sigma_\tau^z|\phi_0}&=\braket{\chi_0^*|V^\dag\sigma_\tau^z V|\phi_0^*}=
-\braket{\chi_0|\sigma_{T-\tau}^z |\phi_0}^*
}
}

 {
Then, we find that $\mr{Re}[\braket{\psi_f|U^{T-\tau}\sigma_i^zU^{\tau}|\psi_i}]=-\mr{Re}[\braket{\psi_f|U^{\tau}\sigma_i^zU^{T-\tau}|\psi_i}]$ if the symmetry is unbroken [the spin indices $i$ is arbitrary].
On the other hand, we can also find $\mr{Re}[\braket{\psi_f|U^{T-\tau}\sigma_i^zU^{\tau}|\psi_i}]\neq-\mr{Re}[\braket{\psi_f|U^{\tau}\sigma_i^zU^{T-\tau}|\psi_i}]$ for the symmetry-breaking phase.
}

\begin{figure}
\begin{center}
\includegraphics[width=\linewidth]{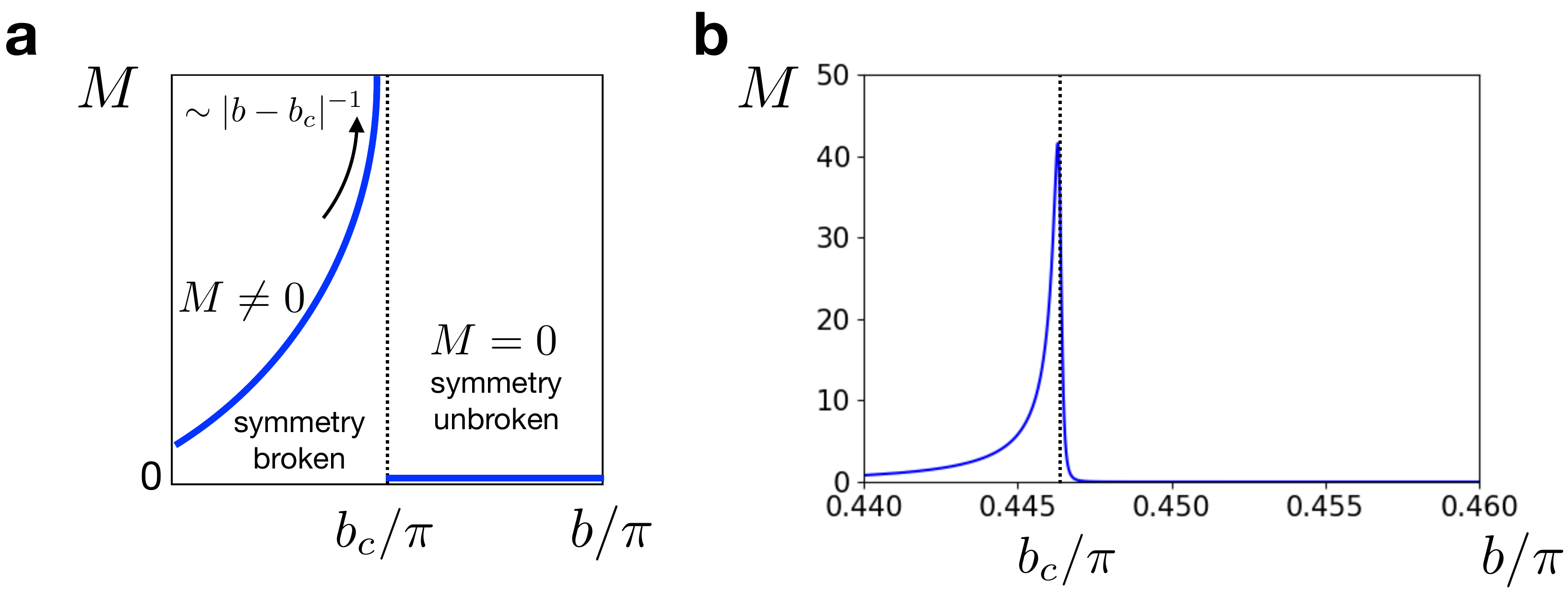}
\end{center}
\caption{Supplementary Figure 8.
\textbf{The order parameter in Supplementary Equation~\eqref{tc}. }
\textbf{a,}  Schematic illustration of the behavior for $M$ for infinite system size.
\textbf{b,}  Numerical verification. The exceptional dynamical quantum phase transition for $F^{\downarrow\uparrow}_{L,T}$ occurs for $b=b_c \simeq 0.446\pi$, where $b<b_c$ represents the symmetry-broken and $b>b_c$ represents the symmetry-unbroken phase.
We find $M\neq 0$ for $b<b_c$ and $M\simeq 0$ for $b>b_c$ (the deviation from zero and the finite peak are due to the finite symmetry-breaking term, which is required for finite $L$).
We use $L=800$, $T=5$, $J=-0.2498\pi$ and $h=1.3$.
}
\label{fig1}
\end{figure}

 {
Then, we can explicitly construct the following order parameter using generalized observables:
\aln{\label{tc}
M=\lrv{\mr{Re}\lrl{\sum_{\tau=1}^{T-1}\frac{\braket{\psi_f|U^{T-\tau}\sigma_i^zU^{\tau}|\psi_i}}{\braket{\psi_f|U^T|\psi_i}}}}^2.
}
}

 {
When the antiunitary symmetry is unbroken, $M=0$, and when it is broken, $M\neq 0$ and behaves as $\sim |b-b_c|^{-1}$ at criticality (Supplementary Figure~\ref{fig1}\textbf{a}).
As shown in Supplementary Figure~\ref{fig1}\textbf{b}, we can demonstrate this using a numerical simulation. 
Here, the slight deviation from $M=0$ for the symmetry-unbroken phase and the finite peak at the critical point are due to the finite symmetry-breaking term, which is required for finite $L$ to demonstrate the symmetry breaking.
}

{
We note that $M$ is constructed using the different-time generalized observables rather than a single-time expectation value of local observables. 
In the above example, $M$ diagnoses  symmetry of interference structure in quantum dynamics related to the time-reversal operation $\tau \leftrightarrow T-\tau$.
}

\section*{Supplementary Note 8: Other models that exhibit exceptional DQPT}
\subsection*{Floquet unitary circuits}
\begin{figure}
\begin{center}
\includegraphics[width=12cm]{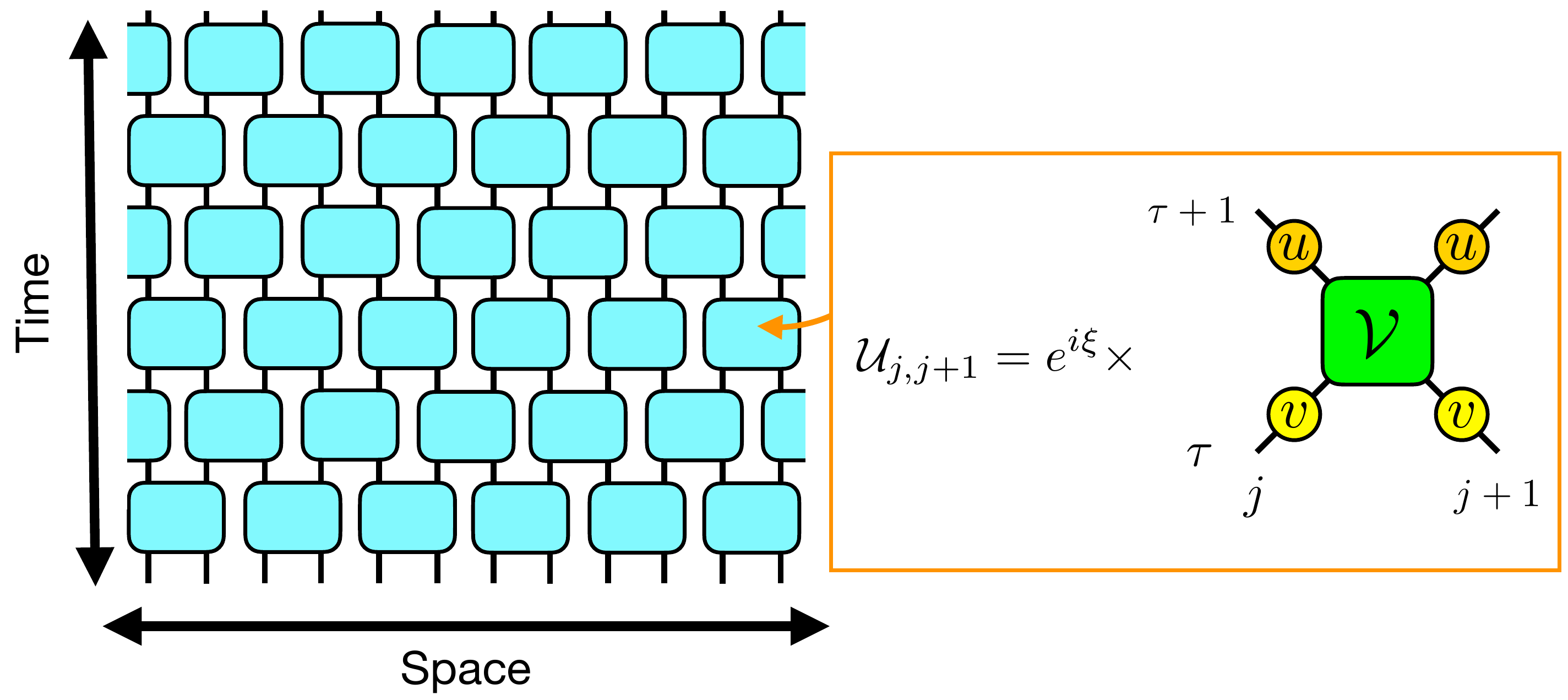}
\end{center}
\caption{Supplementary Figure 9.
\textbf{Schematic illustration of the Floquet circuit dynamics in Supplementary Equation~\eqref{fmodel}.}
We show the example for $L=12$ and $T=6$.
Each of the block $\mc{U}_{j,j+1}$ can be written as in Supplementary Equation~\eqref{block}.
}
\label{circuit}
\end{figure}

Here, we describe in detail that exceptional DQPT can occur in certain Floquet unitary circuits, in addition to our stroboscopic Ising model.
We assume that the system size $L$ is even and consider the unitary circuit given in the form as
\aln{\label{fmodel}
U=\prod_{j\mr{:even}}\mc{U}_{j,j+1}\prod_{j\mr{:odd}}\mc{U}_{j,j+1},
}
which is composed of two-site unitary circuits $\mc{U}_{j,j+1}$ (Supplementary Figure~\ref{circuit}).
When the dimension of the local Hilbert space is two (i.e., spin-1/2 systems), such two-site unitary circuits can be generally represented as~\cite{Bertini19EC}
\aln{\label{block}
\mc{U}_{j,j+1}=e^{i\xi}(u_{j}\otimes u_{j+1})\mc{V}(v_{j}\otimes v_{j+1}),
}
where $u_j$ and $v_j$ are one-site unitary operators, $\xi\in\mbb{R}$, and $\mc{V}$ can be parametrized as
\aln{
\mc{V}=e^{-i(J_{1} \sigma^{x}_j\sigma^{x}_{j+1}+J_{2} \sigma^{y}_j \sigma^{y}_{j+1}+J_{3} \sigma^{z}_j \sigma^{z}_{j+1})}
}
using $J_1,J_2,J_3\in\mbb{R}$.
For simplicity, we assume that $u_j=u$ and $v_j=v$ are $j$-independent in the following.

We now focus on time evolution for $T/2$ steps, which correspond to the total time $T\:(\in 2\mbb{N})$, 
and the following (real part of) dynamical free energy:
\aln{
F^\mr{Tr}_{L,T/2}=-\frac{1}{L}\log|\Tr[U^{T/2}]|+\log 2.
}

We require that the spacetime-dual operator  $\tilde{U}$ of $U$ should satisfy
\aln{
F^\mr{Tr}_{L,T/2}=-\frac{1}{L}\log|\Tr[\tilde{U}^{L/2}]|,
}
with which $F^\mr{Tr}_{\infty,T/2}$ is given by $-(\log |\lambda_\mr{M}|)/2$.

By considering the dual operators for $\mc{U}_{j,j+1}$~\cite{Bertini19EC}, we find
\aln{
\tilde{U}=\frac{1}{4}\prod_{\tau\mr{:even}}\tilde{\mc{U}}_{\tau,\tau+1}\prod_{\tau\mr{:odd}}\mc{\tilde{U}}_{\tau,\tau+1},
}
where
\aln{
\tilde{\mc{U}}_{\tau,\tau+1}=e^{i\xi}(v^\mr{T}\otimes u)\tilde{\mc{V}}_{\tau,\tau+1}
(v\otimes u^\mr{T}).
}
Here, 
\aln{
\tilde{\mc{V}}_{\tau,\tau+1}=\frac{1}{2}e^{-iJ_3+iJ_-}\sigma^z_\tau\sigma^z_{\tau+1}
+\frac{1}{2}e^{-iJ_3-iJ_-}
+\frac{1}{2}e^{iJ_3-iJ_+}\sigma^x_\tau\sigma^x_{\tau+1}
-\frac{1}{2}e^{iJ_3+iJ_+}\sigma^y_\tau\sigma^y_{\tau+1}
}
with $J_\pm=J_1\pm J_2$ and $v^\mr{T}$ denotes the transposition of $v$.
It can also be written as~\cite{Bertini19EC}
\aln{\label{rep}
\left[\begin{array}{cccc}e^{-i J_{3}} \cos \left(J_{-}\right) & 0 & 0 & e^{i J_{3}} \cos \left(J_{+}\right) \\ 0 & -i e^{-i J_{3}} \sin \left(J_{-}\right) & -i e^{i J_{3}} \sin \left(J_{+}\right) & 0 \\ 0 & -i e^{i J_{3}} \sin \left(J_{+}\right) & -i e^{-i J_{3}} \sin \left(J_{-}\right) & 0 \\ e^{i J_{3}} \cos \left(J_{+}\right) & 0 & 0 & e^{-i J_{3}} \cos \left(J_{-}\right)\end{array}\right].
}

\subsection*{Antiunitary symmetry}
To discuss the existence of the exceptional DQPT, we restrict ourselves to $\xi=0$ and 
\aln{
u=v=e^{-i\frac{h}{2}\sigma^z},
}
which correspond to the presence of the uniform magnetic field.
To simplify the notation, we consider the unitary transformation for $\tilde{U}\ra (u\otimes u)\tilde{U}(u^\dag\otimes u^\dag)$, which does not change its eigenvalues, and discuss
\aln{
\tilde{U}=\frac{1}{4}\prod_{\tau\mr{:even}}e^{-ih\sigma^z_\tau} e^{-ih\sigma^z_{\tau+1}}\tilde{\mc{V}}_{\tau,\tau+1}\prod_{\tau\mr{:odd}}e^{-ih\sigma^z_\tau} e^{-ih\sigma^z_{\tau+1}}\mc{\tilde{V}}_{\tau,\tau+1}
}
In the following, we show that $\tilde{U}$ has the AUS and that  the exceptional DQPT  can exist for nontrivial points $J_3=\pi/4$ and $J_3=\pi/2$.

Let us first consider the case for $J_3=\pi/2$.
In this case, we show that
\aln{
V=\prod_{\tau=1}^Te^{i\frac{\pi}{2}\sigma_\tau^x}=i^T\prod_{\tau=1}^T\sigma_\tau^x
}
becomes the AUS.
To see this, we  note that
\aln{
V{\tilde{U}}V^\dag=\frac{1}{4}\prod_\mr{\tau:even}V_2 e^{-ih(\sigma^z_{\tau}+\sigma^z_{\tau+1})}\tilde{\mc{V}}_{\tau,\tau+1}V_2^\dag\prod_\mr{\tau:odd}V_2 e^{-ih(\sigma^z_{\tau}+\sigma^z_{\tau+1})}\tilde{\mc{V}}_{\tau,\tau+1}V_2^\dag,
}
where $V_2$ is a shorthand notation for $\sigma_\tau^x\sigma_{\tau+1}^x$.
First, nonzero matrix elements of local gates satisfy
\aln{
\braket{\uparrow\downarrow|V_2e^{-ih(\sigma^z_{\tau}+\sigma^z_{\tau+1})}\tilde{\mc{V}}_{\tau,\tau+1}V_2^\dag|\uparrow\downarrow}
&=\braket{\downarrow\uparrow|e^{-ih(\sigma^z_{\tau}+\sigma^z_{\tau+1})}\tilde{\mc{V}}_{\tau,\tau+1}|\downarrow\uparrow}\nonumber\\
&=\braket{\uparrow\downarrow|e^{-ih(\sigma^z_{\tau}+\sigma^z_{\tau+1})}\tilde{\mc{V}}_{\tau,\tau+1}|\uparrow\downarrow}^*=-\sin J_-\\
\braket{\downarrow\uparrow|V_2e^{-ih(\sigma^z_{\tau}+\sigma^z_{\tau+1})}\tilde{\mc{V}}_{\tau,\tau+1}V_2^\dag|\uparrow\downarrow}
&=\braket{\uparrow\downarrow|e^{-ih(\sigma^z_{\tau}+\sigma^z_{\tau+1})}\tilde{\mc{V}}_{\tau,\tau+1}|\downarrow\uparrow}\nonumber\\
&=\braket{\downarrow\uparrow|e^{-ih(\sigma^z_{\tau}+\sigma^z_{\tau+1})}\tilde{\mc{V}}_{\tau,\tau+1}|\uparrow\downarrow}^*=\sin J_+\\
\braket{\downarrow\downarrow|V_2e^{-ih(\sigma^z_{\tau}+\sigma^z_{\tau+1})}\tilde{\mc{V}}_{\tau,\tau+1}V_2^\dag|\uparrow\uparrow}
&=\braket{\uparrow\uparrow|e^{-ih(\sigma^z_{\tau}+\sigma^z_{\tau+1})}\tilde{\mc{V}}_{\tau,\tau+1}|\downarrow\downarrow}\nonumber\\
&=-\braket{\downarrow\downarrow|e^{-ih(\sigma^z_{\tau}+\sigma^z_{\tau+1})}\tilde{\mc{V}}_{\tau,\tau+1}|\uparrow\uparrow}^*=i\cos J_+\\
\braket{\downarrow\downarrow|V_2e^{-ih(\sigma^z_{\tau}+\sigma^z_{\tau+1})}\tilde{\mc{V}}_{\tau,\tau+1}V_2^\dag|\downarrow\downarrow}
&=\braket{\uparrow\uparrow|e^{-ih(\sigma^z_{\tau}+\sigma^z_{\tau+1})}\tilde{\mc{V}}_{\tau,\tau+1}|\uparrow\uparrow}\nonumber\\
&=-\braket{\downarrow\downarrow|e^{-ih(\sigma^z_{\tau}+\sigma^z_{\tau+1})}\tilde{\mc{V}}_{\tau,\tau+1}|\downarrow\downarrow}^*=-ie^{-2ih}\cos J_-\\
\braket{\uparrow\uparrow|V_2e^{-ih(\sigma^z_{\tau}+\sigma^z_{\tau+1})}\tilde{\mc{V}}_{\tau,\tau+1}V_2^\dag|\uparrow\uparrow}
&=\braket{\downarrow\downarrow|e^{-ih(\sigma^z_{\tau}+\sigma^z_{\tau+1})}\tilde{\mc{V}}_{\tau,\tau+1}|\downarrow\downarrow}\nonumber\\
&=-\braket{\uparrow\uparrow|e^{-ih(\sigma^z_{\tau}+\sigma^z_{\tau+1})}\tilde{\mc{V}}_{\tau,\tau+1}|\uparrow\uparrow}^*=-ie^{2ih}\cos J_-
}
and the other matrix elements are zero, where we have used Supplementary Equation~\eqref{rep} with $J_3=\pi/2$.
From this, 
while matrix elements of the two-site transitions for
$\{\ket{\uparrow\downarrow}\ra \ket{\uparrow\downarrow}, \ket{\uparrow\downarrow}\ra \ket{\downarrow\uparrow}, \ket{\downarrow\uparrow}\ra \ket{\downarrow\uparrow}, \ket{\downarrow\uparrow}\ra \ket{\uparrow\downarrow}\}$ (yellow gates in Supplementary Figure~\ref{proof})
are invariant under complex conjugation, those of
the two-site transitions for $\lrm{\ket{\uparrow\uparrow}\ra \ket{\uparrow\uparrow}, \ket{\uparrow\uparrow}\ra \ket{\downarrow\downarrow}, \ket{\downarrow\downarrow}\ra \ket{\uparrow\uparrow}, \ket{\downarrow\downarrow}\ra \ket{\downarrow\downarrow}}$ (green gates in Supplementary Figure~\ref{proof}) acquire a minus sign under complex conjugation.

Now, consider matrix elements of ${\tilde{U}}$ as a sum of the paths of the computational states. 
For example, we can consider a matrix element 
\aln{
\braket{\uparrow\uparrow\downarrow\downarrow\downarrow\uparrow|{\tilde{U}}|\uparrow\downarrow\uparrow\uparrow\downarrow\downarrow}=
\frac{1}{4}\sum_{\mbf{s}}&
\braket{\uparrow\uparrow\downarrow\downarrow\downarrow\uparrow|\prod_\mr{\tau:even}{e^{-ih(\sigma^z_{\tau}+\sigma^z_{\tau+1})}\tilde{\mc{V}}_{\tau,\tau+1}}|\mbf{s}}\nonumber\\
&\times\braket{\mbf{s}|\prod_\mr{\tau:odd}{e^{-ih(\sigma^z_{\tau}+\sigma^z_{\tau+1})}\tilde{\mc{V}}_{\tau,\tau+1}}|\uparrow\downarrow\uparrow\uparrow\downarrow\downarrow}
}
 for $T=6$.
 In Supplementary Figure~\ref{proof}, we show one of the paths that corresponds to $\ket{\mbf{s}}=\ket{\uparrow\downarrow\uparrow\uparrow\uparrow\uparrow}$ for this example.
Then, generally, we can show that each of the paths must include even times of two-site transitions for green gates, and even times for yellow gates.
To see this, we focus on the difference of magnetization $\delta m$ between even and odd sites (in the time direction).
For the above example, $\delta m$ is $-1$ for $\ket{\uparrow\downarrow\uparrow\uparrow\downarrow\downarrow}$ and $\ket{\uparrow\downarrow\uparrow\uparrow\uparrow\uparrow}$, and $+1$ for  $\ket{\uparrow\uparrow\downarrow\downarrow\downarrow\uparrow}$, where we assign $-\frac{1}{2}$ for $\ket{\downarrow}$ and $\frac{1}{2}$ for $\ket{\uparrow}$.
When we consider a general matrix element $\braket{\psi_2|\tilde{U}|\psi_1}\propto\sum_\mbf{s}\braket{\psi_2|\cdots|\mbf{s}}\braket{\mbf{s}|\cdots|\psi_1}$, $\delta m$ for $\ket{\psi_1},\ket{\psi_2}, \ket{\mbf{s}}$ have the same even-odd parity.
It is also clear that there are odd/even numbers  of yellow gates for odd/even $\delta m$ in the first half of the path ($\braket{\mbf{s}|\cdots|\psi_1}$) and 
for odd/even $-\delta m$ in the second half of the path ($\braket{\psi_2|\cdots|\mbf{s}}$).
Since $\delta m$ and $-\delta m$  have the same even-odd parity, the total number of the yellow gates in the path is even.
Because the total number of all gates is even, the number of green gates is also even.

\begin{figure}
\begin{center}
\includegraphics[width=12cm]{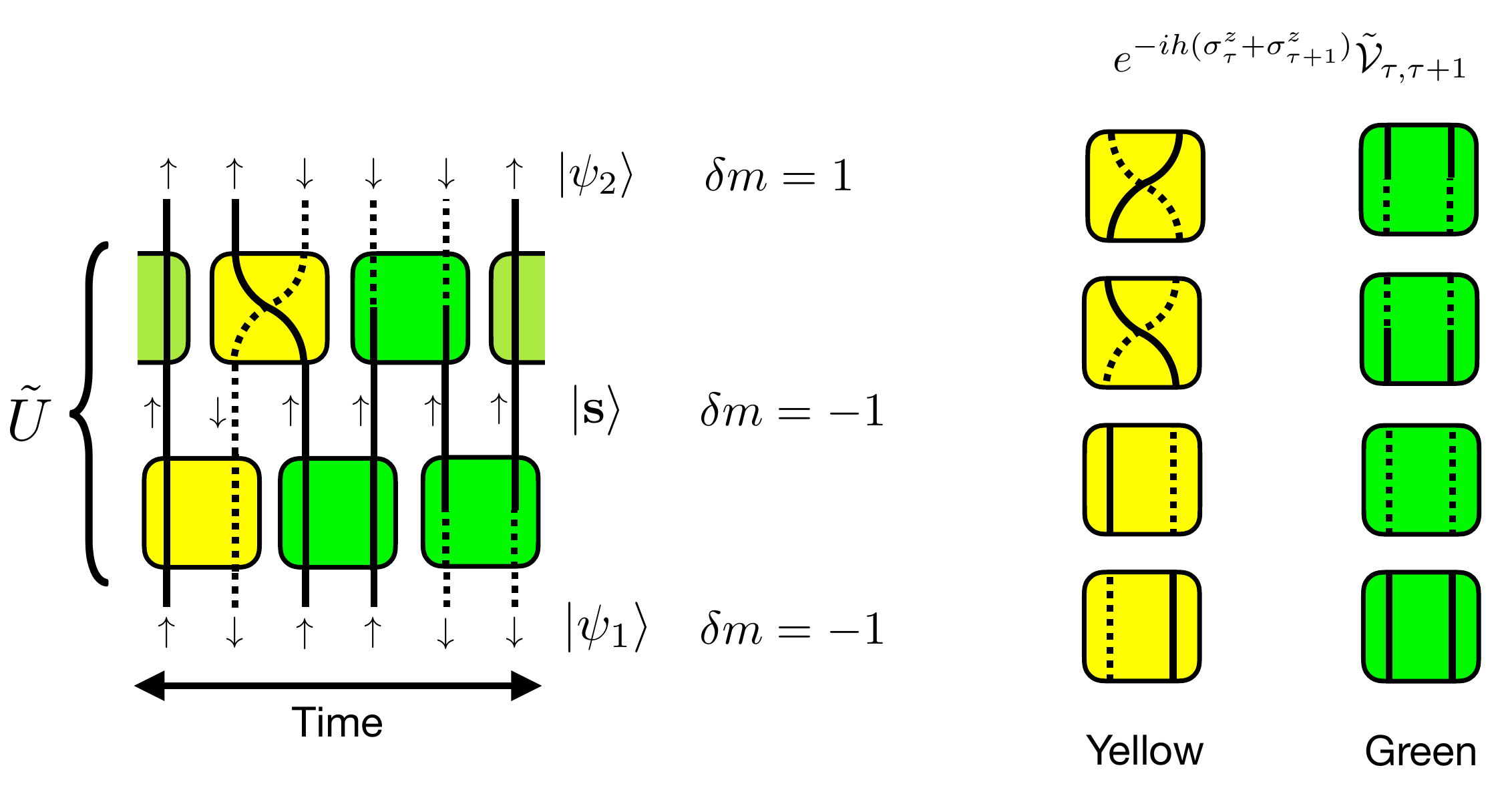}
\end{center}
\caption{Supplementary Figure 10.
\textbf{Example of a path of computational basis that constitutes a matrix element 
$\braket{\psi_2|\tilde{U}|\psi_1}$ of the spacetime-dual operator $\tilde{U}$.}
We here consider $\ket{\psi_1}=\ket{\uparrow\downarrow\uparrow\uparrow\downarrow\downarrow}$ and $\ket{\psi_2}=\ket{\uparrow\uparrow\downarrow\downarrow\downarrow\uparrow}$, and insert a middle state $\ket{\mbf{s}}=\ket{\uparrow\downarrow\uparrow\uparrow\uparrow\uparrow}$, where black lines denote $\ket{\uparrow}$  and dotted lines denote $\ket{\downarrow}$.
These states are transformed by local two-site gates $e^{-ih(\sigma^z_{\tau}+\sigma^z_{\tau+1})}\tilde{\mc{V}}_{\tau,\tau+1}$, which are colored with yellow or green depending on the transformed spin states.
By considering the difference of magnetization $\delta m$ between even and odd sites, we can show that green and yellow gates must appear even times.
}
\label{proof}
\end{figure}

Owing to the even-time appearance of the green gates, the $-1$ phase under complex conjugation for the green gates cancels out for every path.
Then we finally have
\aln{
\braket{\psi_2|V{\tilde{U}}V^\dag|\psi_1}=
\braket{\psi_2|{\tilde{U}}|\psi_1}^*
}
for every matrix element,
i.e., $\tilde{U}$ has the AUS.
We note that, since $V^2=1$ for all $T$,  $\tilde{U}$ belongs to symmetry Class AI  and can have an exceptional point irrespective of $T$, in contrast with the case for the stroboscopic Ising model.

The symmetry structure for $J_3=\pi/4$ can be discussed similarly: we find that 
\aln{
V=\prod_{\tau:\mr{odd}}e^{i\frac{\pi}{2}\sigma_\tau^x}\prod_{\tau:\mr{even}}e^{i\frac{\pi}{2}\sigma_\tau^y}=i^T\prod_{\tau:\mr{odd}}\sigma_\tau^x\prod_{\tau:\mr{even}}\sigma_\tau^y
}
becomes AUS in this case.
To see this, we  note that
\aln{
V{\tilde{U}}V^\dag=\frac{1}{4}\prod_\mr{\tau:even}V'_2 e^{-ih(\sigma^z_{\tau}+\sigma^z_{\tau+1})}\tilde{\mc{V}}_{\tau,\tau+1}{V_2'}^\dag\prod_\mr{\tau:odd}V_2 e^{-ih(\sigma^z_{\tau}+\sigma^z_{\tau+1})}\tilde{\mc{V}}_{\tau,\tau+1}V_2^\dag,
}
where $V_2$ and $V'_2$ are shorthand notations for $\sigma_\tau^x\sigma_{\tau+1}^y$ and $\sigma_\tau^y\sigma_{\tau+1}^x$, respectively.
Again, we can calculate nonzero matrix elements of local gates as
\aln{
\braket{\uparrow\downarrow|V_2e^{-ih(\sigma^z_{\tau}+\sigma^z_{\tau+1})}\tilde{\mc{V}}_{\tau,\tau+1}V_2^\dag|\uparrow\downarrow}
&=\braket{\downarrow\uparrow|e^{-ih(\sigma^z_{\tau}+\sigma^z_{\tau+1})}\tilde{\mc{V}}_{\tau,\tau+1}|\downarrow\uparrow}\nonumber\\
&=i\braket{\uparrow\downarrow|e^{-ih(\sigma^z_{\tau}+\sigma^z_{\tau+1})}\tilde{\mc{V}}_{\tau,\tau+1}|\uparrow\downarrow}^*=-ie^{-\frac{\pi}{4}i}\sin J_-\\
\braket{\downarrow\uparrow|V_2e^{-ih(\sigma^z_{\tau}+\sigma^z_{\tau+1})}\tilde{\mc{V}}_{\tau,\tau+1}V_2^\dag|\uparrow\downarrow}
&=-\braket{\uparrow\downarrow|e^{-ih(\sigma^z_{\tau}+\sigma^z_{\tau+1})}\tilde{\mc{V}}_{\tau,\tau+1}|\downarrow\uparrow}\nonumber\\
&=i\braket{\downarrow\uparrow|e^{-ih(\sigma^z_{\tau}+\sigma^z_{\tau+1})}\tilde{\mc{V}}_{\tau,\tau+1}|\uparrow\downarrow}^*=-e^{-\frac{\pi}{4}i}\sin J_+\\
\braket{\downarrow\downarrow|V_2e^{-ih(\sigma^z_{\tau}+\sigma^z_{\tau+1})}\tilde{\mc{V}}_{\tau,\tau+1}V_2^\dag|\uparrow\uparrow}
&=-\braket{\uparrow\uparrow|e^{-ih(\sigma^z_{\tau}+\sigma^z_{\tau+1})}\tilde{\mc{V}}_{\tau,\tau+1}|\downarrow\downarrow}\nonumber\\
&=-i\braket{\downarrow\downarrow|e^{-ih(\sigma^z_{\tau}+\sigma^z_{\tau+1})}\tilde{\mc{V}}_{\tau,\tau+1}|\uparrow\uparrow}^*=-ie^{-\frac{\pi}{4}i}\cos J_+\\
\braket{\downarrow\downarrow|V_2e^{-ih(\sigma^z_{\tau}+\sigma^z_{\tau+1})}\tilde{\mc{V}}_{\tau,\tau+1}V_2^\dag|\downarrow\downarrow}
&=\braket{\uparrow\uparrow|e^{-ih(\sigma^z_{\tau}+\sigma^z_{\tau+1})}\tilde{\mc{V}}_{\tau,\tau+1}|\uparrow\uparrow}\nonumber\\
&=-i\braket{\downarrow\downarrow|e^{-ih(\sigma^z_{\tau}+\sigma^z_{\tau+1})}\tilde{\mc{V}}_{\tau,\tau+1}|\downarrow\downarrow}^*=e^{-\frac{\pi}{4}i}e^{-2ih}\cos J_-\\
\braket{\uparrow\uparrow|V_2e^{-ih(\sigma^z_{\tau}+\sigma^z_{\tau+1})}\tilde{\mc{V}}_{\tau,\tau+1}V_2^\dag|\uparrow\uparrow}
&=\braket{\downarrow\downarrow|e^{-ih(\sigma^z_{\tau}+\sigma^z_{\tau+1})}\tilde{\mc{V}}_{\tau,\tau+1}|\downarrow\downarrow}\nonumber\\
&=-i\braket{\uparrow\uparrow|e^{-ih(\sigma^z_{\tau}+\sigma^z_{\tau+1})}\tilde{\mc{V}}_{\tau,\tau+1}|\uparrow\uparrow}^*=e^{-\frac{\pi}{4}i}e^{2ih}\cos J_-,
}
where $J_3=\pi/4$ is used.
Again, the two-site transitions for $\{\ket{\uparrow\uparrow}\ra \ket{\uparrow\uparrow}, \ket{\uparrow\uparrow}\ra \ket{\downarrow\downarrow},$ $\ket{\downarrow\downarrow}\ra \ket{\uparrow\uparrow}, \ket{\downarrow\downarrow}\ra \ket{\downarrow\downarrow}\}$ appear even times.
Consequently, the complex conjugation operation leaves the overall factor $i^T$, i.e.,
\aln{
\braket{\psi_2|V{\tilde{U}}V^\dag|\psi_1}=
(-1)^{T/2}\braket{\psi_2|{\tilde{U}}|\psi_1}^*,
}
meaning that $\tilde{U}$ has the AUS.
We note that $VV^*=(-1)^{T/2}$ and thus $\tilde{U}$ belongs to Class AI/AII for even/odd $T/2$.
Thus, the exceptional DQPT occurs only when $T/2$ is even in the case of $J_3=\pi/4$.

\subsection*{Dynamical phase transitions}
We demonstrate that the exceptional DQPT occurs for the above Floquet circuit model.
In Supplementary Figure~\ref{modelchange}, we show dynamical free energy $F^\mr{Tr}_{\infty,T/2}$ for $J_3=\pi/2$ and $J_3=\pi/4$, where we vary $J_2$.
We find that the exceptional DQPTs occur for both of the cases, thanks to the antiunitary symmetry hidden in the spacetime-dual operator $\tilde{U}$.
We also note that, there appear DQPTs through the self-dual points~\cite{Bertini19EC} with $J_3=\pi/4$ and $J_2=\pi/4,3\pi/4$, where $F^\mr{Tr}_{\infty,T/2}$ universally takes $\log 2$.

\begin{figure}
\begin{center}
\includegraphics[width=\linewidth]{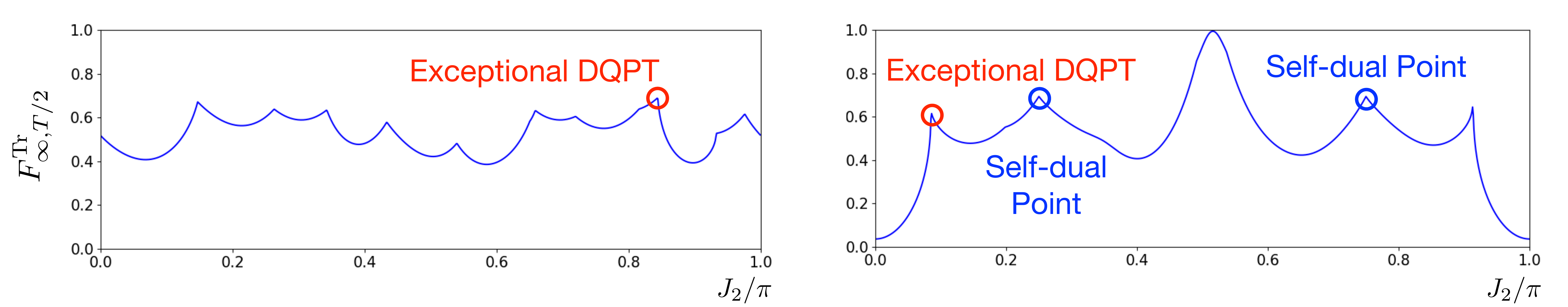}
\end{center}
\caption{Supplementary Figure 11.
\textbf{(Real part of) dynamical free energy $F^\mr{Tr}_{\infty,T/2}$ as a function of $J_2$.}
We choose parameters as (left) $J_3=\pi/2, J_1=0.15\pi, h=1.0, T=6$ and
(right) $J_3=\pi/4, J_1=0.47\pi, h=1.0, T=8$.
As varying $J_2$, the exceptional dynamical quantum phase transition (DQPT) occurs for both of the cases, as well as DQPTs through self-dual points for $J_3=\pi/4$.
}
\label{modelchange}
\end{figure}

%\paragraph{Implication to classical dynamical phase transitions}

\bibliographystyle{naturemag}
\bibliography{../refer_them}